\documentclass[iop,apj]{emulateapj}
\usepackage{amsmath,amssymb}

\usepackage{xcolor}

\slugcomment{The Astrophysical Journal, 769:24 (10pp), 2013 May 20}

\usepackage[bookmarksnumbered,
colorlinks,
linkcolor=blue,
citecolor=black,
filecolor=black,
urlcolor=blue,
breaklinks=true,
]{hyperref}
\shortauthors{Knobel et al.}
\begin{document}
\title{The Colors of Central and Satellite galaxies in zCOSMOS out to $z \simeq 0.8$ and implications for quenching\footnotemark[1]}

\author{C.~Knobel\footnotemark[2],
S.~J.~Lilly\footnotemark[2],
K.~Kova\v{c}\footnotemark[2],
Y.~Peng\footnotemark[2],
T.~J.~Bschorr\footnotemark[2],
%
C.~M.~Carollo\footnotemark[2],
T.~Contini\footnotemark[3,4],
J.-P.~Kneib\footnotemark[5],
O.~Le Fevre\footnotemark[5],
V.~Mainieri\footnotemark[6],
A.~Renzini\footnotemark[7],
M.~Scodeggio\footnotemark[8],
G.~Zamorani\footnotemark[9],
%
S.~Bardelli\footnotemark[9],
M.~Bolzonella\footnotemark[9],
A.~Bongiorno\footnotemark[10],
K.~Caputi\footnotemark[2,20],
O.~Cucciati\footnotemark[11],
S.~de la Torre\footnotemark[12],
L.~de Ravel\footnotemark[12],
P.~Franzetti\footnotemark[8],
B.~Garilli\footnotemark[8],
A.~Iovino\footnotemark[11],
P.~Kampczyk\footnotemark[2],
F.~Lamareille\footnotemark[3,4],
J.-F.~Le Borgne\footnotemark[3,4],
V.~Le Brun\footnotemark[5],
C.~Maier\footnotemark[2,19],
M.~Mignoli\footnotemark[9],
R.~Pello\footnotemark[3,4],
E.~Perez Montero\footnotemark[3,4,13],
V.~Presotto\footnotemark[11,22],
J.~Silverman\footnotemark[23],
M.~Tanaka\footnotemark[14],
L.~Tasca\footnotemark[5],
L.~Tresse\footnotemark[5],
D.~Vergani\footnotemark[9,21],
E.~Zucca\footnotemark[9],
%
%
L.~Barnes\footnotemark[2],
R.~Bordoloi\footnotemark[2],
A.~Cappi\footnotemark[9],
A.~Cimatti\footnotemark[15],
G.~Coppa\footnotemark[10],
A.~M.~Koekemoer\footnotemark[16],
C.~L\'opez-Sanjuan\footnotemark[5,24],
H.~J.~McCracken\footnotemark[17],
M.~Moresco\footnotemark[15],
P.~Nair\footnotemark[9],
L.~Pozzetti\footnotemark[9],
and N.~Welikala\footnotemark[18]
}

\footnotetext[1]{European Southern Observatory (ESO), Large Program 175.A-0839}

\affil{$^2$Institute for Astronomy, ETH Zurich, Zurich 8093, Switzerland\\
$^3$Institut de Recherche en Astrophysique et Plan\'etologie, CNRS, 14, avenue Edouard Belin, F-31400 Toulouse, France\\
$^4$IRAP, Universit\'e de Toulouse, UPS-OMP, Toulouse, France\\
$^5$Laboratoire d'Astrophysique de Marseille, CNRS/Aix-Marseille Universit\'e, 38 rue Fr\'ed\'eric Joliot-Curie, F-13388 Marseille cedex 13, France\\
$^6$European Southern Observatory, Garching, Germany\\
$^7$INAF-Osservatorio Astronomico di Padova, Vicolo dell'Osservatorio 5, 35122, Padova, Italy\\
$^8$INAF-IASF Milano, Milano, Italy\\
$^9$INAF Osservatorio Astronomico di Bologna, via Ranzani 1, I-40127, Bologna, Italy\\
$^{10}$Max Planck Institut f\"ur Extraterrestrische Physik, Garching, Germany\\
$^{11}$INAF Osservatorio Astronomico di Brera, Milan, Italy\\
$^{12}$Institute for Astronomy, University of Edinburgh, Royal Observatory, Edinburgh, EH93HJ, UK\\
$^{13}$Instituto de Astrofisica de Andalucia, CSIC, Apartado de correos 3004, 18080 Granada, Spain\\
$^{14}$Institute for the Physics and Mathematics of the Universe (IPMU), University of Tokyo, Kashiwanoha 5-1-5, Kashiwa, Chiba 277-8568, Japan\\
$^{15}$Dipartimento di Astronomia, Universit\`a degli Studi di Bologna, Bologna, Italy\\
$^{16}$Space Telescope Science Institute, Baltimore, MD 21218, USA\\
$^{17}$Institut d'Astrophysique de Paris, UMR7095 CNRS, Universit\'e Pierre \& Marie Curie, 75014 Paris, France\\
$^{18}$Insitut d'Astrophysique Spatiale, B\^atiment 121, Universit\'e Paris-Sud XI and CNRS, 91405 Orsay Cedex, France\\
$^{19}$University of Vienna, Department of Astronomy, Tuerkenschanzstrasse 17, 1180 Vienna, Austria\\
$^{20}$Kapteyn Astronomical Institute, University of Groningen, P.O.~Box 800, 9700 AV Groningen, The Netherlands\\
$^{21}$INAF-IASF Bologna, Via P.~Gobetti 101, I-40129 Bologna, Italy\\
$^{22}$Dipartimento di Fisica dell'Universit\`a degli Studi di Trieste - Sezione di Astronomia, via Tiepolo 11, 34143, Trieste, Italy\\
$^{23}$Kavli Institute for the Physics and Mathematics of the Universe, Todai Institutes for Advanced Study, the University of Tokyo, Kashiwa, Japan 277-8583 (Kavli IPMU, WPI)\\
$^{24}$Centro de Estudios de F\'{\i}sica del Cosmos de Arag\'on, Plaza San Juan 1, planta 2, 44001 Teruel, Spain
}

\begin{abstract}

We examine the red fraction of central and satellite galaxies in the large zCOSMOS group catalog out to $z \simeq 0.8$ correcting for both the incompleteness in stellar mass and for the less than perfect purities of the central and satellite samples.  We show that, at all masses and at all redshifts, the fraction of satellite galaxies that have been quenched, i.e., are red, is systematically higher than that of centrals, as seen locally in the Sloan Digital Sky Survey (SDSS). The satellite quenching efficiency, which is the probability that a satellite is quenched because it is a satellite rather than a central, is, as locally, independent of stellar mass.  Furthermore, the average value is about 0.5, which is also very similar to that seen in the SDSS.  We also construct the mass functions of blue and red centrals and satellites and show that these broadly follow the predictions of the Peng et al.~analysis of the SDSS groups.  Together, these results indicate that the effect of the group environment in quenching satellite galaxies was very similar when the universe was about a half its present age, as it is today.\\[2mm]
\end{abstract}

\section{Introduction}
\setcounter{footnote}{24}

One of the most striking features of the galaxy population is the observed bimodality in the color of galaxies \citep[e.g.,][]{strateva2001}: most galaxies are either blue and star forming, with a star formation rate (SFR) that is closely linked to the existing stellar mass, or red with an SFR that is lower by 1-2 orders of magnitude. This bimodality is observed to be already in place at redshift $z \sim 1$ \citep[e.g.,][]{bell2004,cirasuolo2007,mignoli2009} and a key question in the context of galaxy evolution is to understand the mechanism(s) that lead to the cessation of star-formation as a function of stellar mass, cosmic environment, and cosmic time \citep[see, e.g.,][and references therein]{kauffmann2004,balogh2004,baldry2006,peng2010,presotto2012,cibinel2012,woo2013}.  We refer to this process as ``quenching''.

Although the specific mechanisms of this quenching process are still largely debated, there has been some progress in phenomenologically understanding where and when it happens. There is good evidence that the action of quenching is different for galaxies in different locations with respect to the dark matter halos in which they reside. Broadly speaking, ``central'' galaxies are taken to be the most massive galaxies in a halo and to reside at the potential minimum.  ``Satellite'' galaxies are galaxies that are moving relative to the potential minimum having fallen in to the larger halo.  In this way, all galaxies are either centrals or satellites.  The role of the central/satellite dichotomy in quenching has been discussed in several observational papers \citep[e.g.,][]{vandenbosch2008,kimm2009,woo2013,peng2012}.  \cite{vandenbosch2008} introduced the ``satellite quenching efficiency'', which measures the fraction of blue (previously central) galaxies that are quenched when they become satellite galaxies. They showed that in the low-redshift universe sampled by the Sloan Digital Sky Survey (SDSS), this quantity is independent of the stellar mass of the satellite galaxy. 

\cite{peng2010} showed that the fraction of galaxies that are quenched, i.e., the so-called red fraction,  in SDSS is fully separable in terms of the effects of stellar mass and environment. This means that the fraction of galaxies that survive as blue star-forming galaxies can be written as the product of two functions, one of stellar mass and one of environment.  They argued that this indicated the action of two distinct processes, one, ``mass quenching'', that is linked to the stellar mass of a galaxy but not the environment, and the other, ``environment quenching'', that is linked to its environment but not its mass.  In \cite{peng2012} it was shown that all of the environmental effects in the SDSS population can be ascribed to the satellite galaxies in groups.  The colors of central galaxies are independent of environment, and these galaxies must experience only the mass quenching process.  In contrast, satellite galaxies experience both mass-quenching and environment quenching.  It was shown in  \cite{peng2012} that the probability that a satellite had been environment (or satellite) quenched was independent of its stellar mass, as in \cite{vandenbosch2008}, but also that it depended on the local density of galaxies, which was taken as a proxy for the location of the satellite in the group.  Interestingly, at a given density, the satellite quenching efficiency does not depend on the mass of the parent halo.

In this paper, we aim to investigate the properties of centrals and satellites at much higher redshifts, using the 20k group catalog \citep[][``K12'']{knobel2012} from the spectroscopic zCOSMOS-bright survey \citep{lilly2007}. The zCOSMOS 20k group catalog contains about 1500 groups (including pairs) within the redshift range $0.1 \lesssim z \lesssim 1$ and features very good and well controlled statistics in terms of completeness and purity of the group population. Moreover, in K12 we developed a probabilistic scheme to select subsamples of centrals and satellites which have purities up to 80\%.

The definition of the central galaxy is an important issue.  In most studies, the centrals have been identified by just selecting the brightest or the most massive galaxies within dark matter halos. However, both semi-analytic simulations \citep[e.g.,][]{kitzbichler2007,henriques2012} as well as analysis of observational data in the low-redshift universe \citep{skibba2011} indicate that a non-negligible fraction (i.e., 20\%-30\%) of the galaxies that lie at the deepest point within the potential well are not in fact the most luminous or most massive galaxies in their halos. This ambiguity may be exacerbated by uncertainties in the estimation of stellar masses from photometric data. For a discussion of a self-consistent definition of central galaxies we refer to \cite{carollo2012}.

It is likely that the location of the galaxy in the group is more important than the property of being the most massive. For this reason, we developed a probabilistic selection scheme for centrals and satellites in K12 that also includes positional information in addition to stellar mass. This scheme was carefully calibrated using mock catalogs generated from numerical simulations \citep{kitzbichler2007,henriques2012}. These well controlled statistics can be used, in principle, to correct our set of centrals an satellites for misidentifications.

Our goal is to extend the discussion of centrals and satellites up to redshift $z \simeq 0.8$ particularly with focus on the quenching of central and satellite galaxies as a function of stellar mass and cosmic epoch. We therefore study the fraction of centrals and satellites and construct the red fractions of each set. From this we derive the satellite quenching efficiency (averaged over different local densities).  We also look at the mass functions of the blue and the red central and satellite galaxy populations since, as emphasized by \cite{peng2010,peng2012}, this is also a sensitive diagnostic of the quenching action. We discuss our findings in the context of the model proposed by \cite{peng2012}. In a later study (K.~Kovac et al., in preparation), we will analyze the quenching of centrals and satellites in zCOSMOS with emphasis on their dependence on the local environmental density.

The layout of this paper is as follow. In Section \ref{sec:selected_sample}, we describe the data and the samples that we used for our analysis. In Section, \ref{sec:results} we describe our results for the fraction of satellites, the satellite quenching efficiency and the mass functions of the different galaxy populations. Finally, in Section \ref{sec:conclusion} we summarize and conclude our findings.

Where necessary, a concordance cosmology with $H_0 = 70\ \rm{km\; s^{-1}\; \rm{Mpc}^{-1}}$, $\Omega_{\rm m} = 0.25$, and $\Omega_\Lambda = 0.75$ is applied. All magnitudes are quoted in the AB system. We use the term ``dex'' to express the antilogarithm, i.e., 0.1 dex corresponds to a factor $10^{0.1} \simeq 1.259$. By $\log$ we refer to the 10-based logarithm and we occasionally use $\log M$ as short term for $\log(M/M_\odot)$.

\section{Data}\label{sec:selected_sample}

We use the central and satellite classification scheme from the group catalog described in K12, which is based on the $15 \leq I_{\rm AB} \leq 22.5$ flux limited sample from the spectroscopic zCOSMOS-bright survey \citep[][S.~J.~Lilly et al.~2013, in preparation]{lilly2007,lilly2009}. Multiplying the (random) spatial selection with the redshift success rate, this sample has an overall mean sampling rate of the target sample of about 50\%. The catalog was produced by a Friends-of-Friends (FOF) multi-run scheme \citep[see][]{knobel2009} in which successively different group-finding parameters were used, optimized for different richness groups. The halo masses of the groups are typically in the range $12.5 \lesssim \log(M_{\rm h}/M_\odot) \lesssim 13.5$ \citep[cf.][]{knobel2012b} and the non-group galaxies typically populate halos within the mass range $11 \lesssim \log(M_{\rm h}/M_\odot) \lesssim 13$. For details on the construction and the statistics of the group catalog we refer to K12.

As well as attempting to classify all of the galaxies as centrals or satellites, taking into account also those galaxies for which a spectroscopic redshift is not available, a feature of this group catalog is the possibility to also produce fairly pure subsamples of centrals and satellites.  The purity of a (sub)sample is defined as the number of correctly classified galaxies (either centrals or satellites) within the (sub)sample divided by the total number of galaxies in the (sub)sample. In K12 every galaxy (including those only with photometric redshifts, photo-$z$) was assigned a membership probability $p$ to be a member of a given group, a probability $p_{\rm M}$ to be the most massive galaxy within that group, and another parameter $p_{\rm MA}$ which additionally to $p_{\rm M}$ also considers the positional information within the group.  The photo-$z$ galaxies were used in K12 to help classify the spectroscopic galaxies, e.g., by indicating the possible presence of a better candidate for the central, but are not then used in the present analysis because their own purities are naturally much lower.
 
By selecting (spectroscopic) galaxies according to their $p$, $p_{\rm M}$, and $p_{\rm MA}$, (sub)samples of centrals and satellites of varying size and purity can be produced. An important case is the ``dichotomous sample'', in which all galaxies of the zCOSMOS spectroscopic sample are classified to be either a central or a satellite galaxy. Based on simulated mock catalogs, it was found that defining the satellites by simultaneously applying $p > 0.1$, $p_{\rm M} < 0.5$, and $p_{\rm MA} < 0.5$, and identifying all other galaxies (including those that are not associated to any group, i.e., with $p = 0$) as centrals, resulted in a purity (averaged over the whole population) of about 84\% for the centrals and about 74\% for the satellites (see Table 10 of K12).   

It should be appreciated that these purities include both the difficulties of identifying the central of a given group, and the errors that are made in identifying the groups in the first place.  Clearly, a group that has been fragmented into multiple structures, or over-merged with other groups into a spuriously large structure by the group-finder algorithm (see K12 for discussion), will not end up with a 100\% correct classification of centrals and satellites.   

The impurity of the adopted central and satellite populations due to cross-mixing between them, will lead to a reduction of any observed differences in a given property between the two sets.  If the purities are known, then of course this dilution effect can be corrected for, as detailed below.  We would expect the purities to be a function of both redshift and stellar mass, and therefore computed them in this way. The purity functions for the dichotomous sample are shown in Figure \ref{fig:satellite_fraction}. For both centrals and satellites the purity is $\gtrsim 80\%$ at all redshifts and is essentially independent of mass.

Here we assumed that the purity is essentially independent of the color of the galaxy, since color information was neither used for the construction of the group catalog nor for the central/satellite classification. Of course, one could imagine that the variation of impurity from the group center \citep[cf.][Figure 10]{knobel2009} could introduce color dependent impurities due to the radial dependence of quenching from the group center \citep{presotto2012}. However, this effect should be small so that we neglect it.

It should be noted that the purities cited in K12 were obtained using mock catalogs that were derived from the light cones of \cite{kitzbichler2007}, which are in turn based on the Millennium Simulation \citep{springel2005} and the semi-analytic model of \cite{croton2006}. For our present analysis we updated our mock catalogs using the light cones of \cite{henriques2012}, which are also derived from the Millennium Simulation, but are based on the more recent semi-analytic model of \cite{guo2011}. The differences on the derived purities, however, are rather small and our final results are essentially independent of the specific semi-analytic model, from which our mock catalogs were derived.

\subsection{Data Samples}\label{sec:data}

The rest frame absolute magnitudes and stellar masses were derived from spectral energy distribution (SED) fitting to 12 broad band filters using {\tt ZEBRA}+ (see K12 for details). For the stellar masses we used standard \cite{bruzual2003} models with an initial stellar mass function of \cite{chabrier2003} and dust extinction according to \cite{calzetti2000}. The absolute magnitudes were not corrected for dust attenuation. The masses are defined here to be the integral of the star-formation history of the model galaxy, and do not, therefore, include the effects of mass return from the stellar population. The advantage of this choice is, for instance, that red (passive) galaxies with zero SFR have constant mass, enabling straightforward comparisons of stellar mass functions. The uncertainty of the stellar masses is about 0.2 dex. 

There are several ways of dividing the galaxy population into star forming and quiescent galaxies in the literature. The simplest way is by means of a cut in the color-mass diagram \citep[e.g.,][]{baldry2006,vandenbosch2008,peng2010,peng2012}. In this context, it is often pointed out that the usage of a single color is prone to misclassifying dust-reddened star forming galaxies as quiescent, which can be prevented using a color-color diagram instead \citep[e.g.,][]{williams2009,bundy2010,brammer2011,ilbert2010,cheung2012,ilbert2013,mok2013}. Others use spectral/morphological/SED types \citep[e.g.,][]{drory2009,pozzetti2010} or directly use SFR together with stellar mass \citep[e.g.,][]{mcgee2011,wetzel2012,moustakas2013,woo2013}. These different methods can lead to different results and one should be careful in interpreting and comparing the results from different studies. The advantage of color is that it is straightforward to define and in comparing different data sets, but more difficult to interpret. SFR estimates, while more physically relevant, are rather uncertain at high redshift and can be model-dependent depending on the available data.

\begin{figure}
	\centering
	\includegraphics[width=0.48\textwidth]{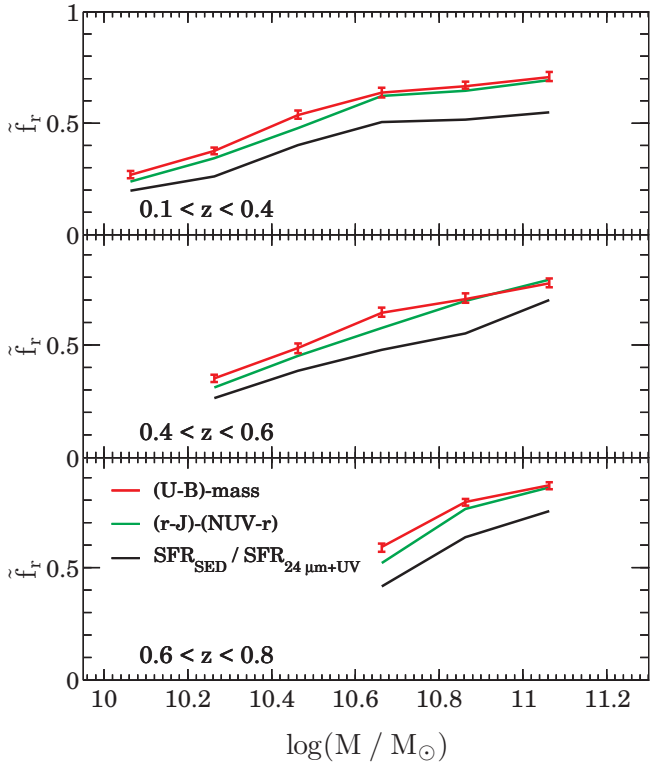}
	\caption{Observed red fraction $\tilde f_{\rm r}$ of the total galaxy population (uncorrected for incompleteness with mass) for different splits into ``blue'' (star forming) and ``red'' (quiescent) galaxies. The red curve corresponds to the split using the $(U-B)$-mass diagram (cf.~Equation (\ref{eq:UB})), the green curve to the split using the $(r-J)$-$(NUV-r)$ color-color diagram, and the black curve to the split using SFR derived from SED fits and derived from 24 $\mu$m and UV data for a subsample of our galaxies (see the text). The error bars (shown only for the red curves) represent the upper and lower quartiles derived from bootstrapping within each mass bin.}\label{fig:red_fraction_comparison}
\end{figure}

To estimate the impact of different ways of splitting the sample on our results, we show in Figure \ref{fig:red_fraction_comparison} the observed red fraction of the total galaxy population for a split using the rest frame $(U-B)$ color-mass diagram, a split using the rest frame $(r-J)$-$(NUV-r)$ color-color diagram, and a split using the SFR-mass diagram set at about 0.5 dex below the median SFR of the red sequence (i.e., we consider galaxies that clearly do not lie on the main sequence as quiescent). The SFR is partially derived from SED fitting and, if available, from 24 $\mu$m and UV data following the method described in \cite{wuyts2011}. The red fractions $\tilde f_{\rm r}$ in Figure \ref{fig:red_fraction_comparison} are not corrected for incompleteness of the different subpopulations with mass --- such corrections will be applied in the subsequent analysis. The red fraction from the color-color diagram is on average only about 6\% lower than the red fraction from the color-mass diagram, while using SFR together with stellar mass leads to a significantly lower red fraction by about 20\% \citep[cf.][]{woo2013}. Interestingly, the systematic relative shifts of the red fraction are only weak functions of mass and redshift. For the details of the derivation of the SFR and for a more detailed discussion of this issue we refer to K.~Kovac et al.~(in preparation).

Since we aim to extend the studies of \cite{vandenbosch2008} and \cite{peng2012} to the high-redshift universe and since the use of color enables us to easily correct for incompleteness of our sample with stellar mass, we will perform our analysis by just appling a simple cut in the $(U-B)$-mass diagram within each redshift bin, where our redshift bins are $0.1 < z < 0.4$, $0.4 < z < 0.6$, and $0.6 < z < 0.8$ (see Table \ref{tab:galaxy_samples}). The color cuts are obtained by eye separating the blue and the red sequences at the site of the lowest number density of objects in the diagram. It turns out that applying the same color cut to each redshift bin works fairly well. In the following, red galaxies are defined by
\begin{equation}\label{eq:UB}
U - B > 1.1 + 0.02\big(\log(M/M_\odot)-10\big)\:,
\end{equation}
where $U$ and $B$ are the rest frame absolute magnitudes and $M$ is the stellar mass. Changing the cuts by 0.02 in color, or making small changes to the slope, has small effects on the results, typically much less than the quoted 1$\sigma$ error bars.
\begin{deluxetable}{crrrcrr}
\tablewidth{0pt}
\tablecaption{Galaxy Samples}
\tablehead{
	\colhead{Redshift} &
	\colhead{$M_{\rm min}$\tablenotemark{a}} & 
 \multicolumn{2}{c}{No.~of Centrals} & 
  \colhead{} &
  \multicolumn{2}{c}{No.~of Satellites} \\
  \cline{3-4} \cline{6-7} \\
	\colhead{} &
	\colhead{} & 
  \colhead{Red} & 
    \colhead{Blue} & 
     \colhead{} &
  \colhead{Red} & 
\colhead{Blue}
  }
\startdata
$0.1 < z < 0.4$ & 9.9 &   788    &     788   &&      343  &       228  \\
$0.4 < z < 0.6$ & 10.1 &  746   &      574   &&     187  &       83  \\
$0.6 < z < 0.8$ & 10.5 &  1122  &       359  &&     212  &        53
\enddata
\tablecomments{The numbers refer to the dichotomous sample (see the text) and consider only galaxies with masses larger than $M_{\rm min}$.}
\tablenotetext{a}{Minimum stellar mass in units of $\log(M_{\rm min}/M_\odot)$.}

\label{tab:galaxy_samples}
\end{deluxetable}

Since our sample is $I$-band flux limited, the completeness of our sample depends on stellar mass and on the redshift. Moreover, the mass-completeness is different for red and blue galaxies.   Given the excellent photo-$z$ in the COSMOS field, we use the photo-$z$ sample down to $I_{\rm AB} < 24$ to empirically determine and correct for this incompleteness.  We apply the same color cut to all COSMOS galaxies and compute for each redshift bin, and for both red and blue galaxies separately, the completeness as a function of stellar mass, i.e., what fraction of galaxies in the photo-$z$ sample are also present in the spectroscopic sample. To compute fractions of galaxies and mass functions, individual objects in the spectroscopic sample are then weighted by the inverse of the completeness. The mass-limit $M_{\rm min}$ of the survey at a given redshift (see Table \ref{tab:galaxy_samples}) is defined to be that mass at which the weighting reaches a factor of two relative to the high mass end.  We checked that our results are not sensitive to the exact choice of $M_{\rm min}$.

This correction method is based on the assumption that, within a given color class (blue or red) the mass-to-light ratio of a galaxy does not depend on whether it is a central or satellite, i.e., on whether it is a group member or not, so that the correction can be applied in the same way to centrals and satellites. This assumption is reasonable as it was shown that the SFR and color distribution for both star forming and quiescent galaxies are essentially independent of environment \citep[e.g.,][]{balogh2004,mcgee2011,peng2012}. Our correction method also naturally deals with the spectroscopic incompleteness of the zCOSMOS sample due to the (inhomogeneous) sampling rate and the redshift success rate (i.e., the fraction of reliably measured redshifts from observed spectra) being a function of selection magnitude, redshift, and color \citep[see Figures 2 and 3]{lilly2009}, as long as the chance of obtaining a reliable redshift does not depend on whether the galaxy is a central or a satellite.

Statistical errors are computed by bootstrapping. That is, to compute the error for a certain quantity (e.g.~a fraction of galaxies or a mass function), we randomly resample our whole galaxy sample 100 times, perform the whole analysis consistently for each of these samples, and finally compute the dispersion (e.g., standard deviation or quartiles) of the corresponding results.

\section{Results}
\label{sec:results}

In this section, we develop the results from our analysis. To make the formalism as simple as possible, we adopt the following conventions. The symbol $N$ denotes number, $f$ fraction, and $P$ purity. For any of these quantities the subscript ``r'' refers to red, ``b'' to blue, ``c'' to central, and ``s'' to satellite, e.g., $f_{\rm s}$ the fraction of satellite galaxies (within a given redshift and mass bin). If subscripts are combined, the first subscript refers to the property that is considered within the sample defined by the second subscript, e.g., $f_{\rm r,c}$ is the fraction of centrals that are red (within a given redshift and mass bin). Whenever a quantity is given a tilde, it is uncorrected with respect to color completeness or central/satellite impurities.

\subsection{Fraction of Satellites}

\begin{figure}
	\centering
	\includegraphics[width=0.48\textwidth]{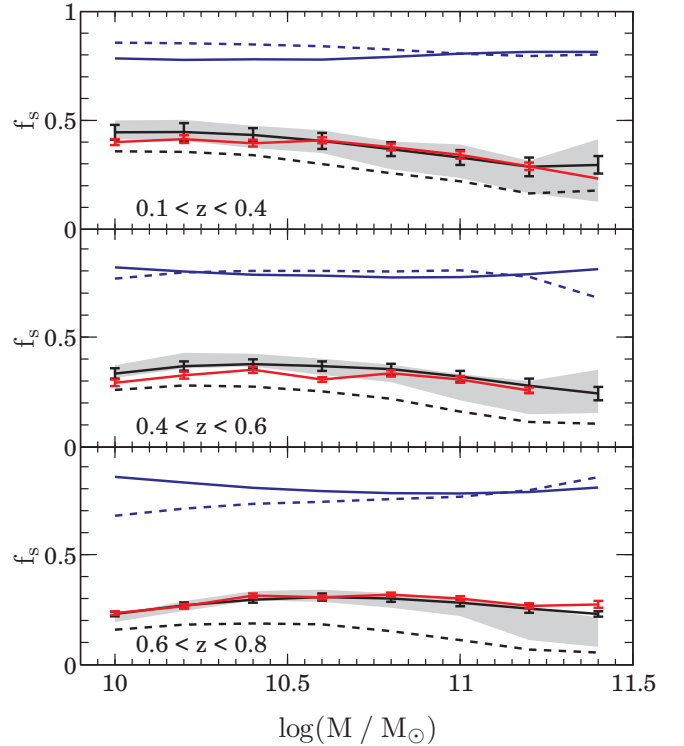}
	\caption{Fraction of satellites $f_{\rm s}$ as a function of stellar mass $M$ within the three redshift bins. The red lines show the (corrected) fraction for the zCOSMOS data with the error bars being the standard deviation derived from bootstrapping. The black lines show the fractions for the corresponding mock samples selected in the same way as the actual data, where the solid lines correspond to the fractions corrected for impurities (see the text) and the dashed lines to the uncorrected fractions. The error bars represent the standard deviation of the 24 mock catalogs. The gray shaded areas represent the $1\sigma$ regions of the fraction of \emph{real} satellites within the 24 mock catalogs. The blue solid lines show the purity of centrals, and the blue dashed lines the purity of satellites. The agreement between the actual data and the mock catalogs is very good.\label{fig:satellite_fraction}}
\end{figure}

Although the purity for centrals and satellites is quite good for all samples (see Figure \ref{fig:satellite_fraction}), the impurities are still on the level of 20\% and this will affect our results. For this reason, we will use these purity functions to correct all of our results, as well as possible, for this contamination. As an example, the dichotomous galaxy sample will consist of $\tilde N_{\rm c}$ galaxies classified as centrals, and $\tilde N_{\rm s}$ as satellites, within a given bin in stellar mass and redshift. If $P_{\rm c}$ is the purity of centrals and $P_{\rm s}$ is the purity of satellites within that bin, the corrected numbers of centrals and satellites are given by
\begin{align}
N_{\rm c} &= \tilde N_{\rm c} - (1-P_{\rm c}) \tilde N_{\rm c} + (1-P_{\rm s}) \tilde N_{\rm s} \label{eq:N_c}\\
N_{\rm s} &= \tilde N_{\rm s} - (1-P_{\rm s}) \tilde N_{\rm s} + (1-P_{\rm c})\tilde N_{\rm c} ,\label{eq:N_s}
\end{align}
respectively.

The fraction of satellites as a function of mass is shown in Figure \ref{fig:satellite_fraction} in the three redshift bins and is compared to the mock catalogs that were derived from the simulated light cones of \cite{henriques2012}. The red lines show the purity-corrected satellite fraction for our actual zCOSMOS data, the black lines show the purity corrected satellite fraction for the corresponding sample in the mock catalogs, and the gray shaded area shows the 1$\sigma$ range for the real fraction of satellites in the mock catalogs. The dashed lines show the fraction of satellites for the mock sample, if we do not correct for impurities, and show the size of effect that is being corrected for. It is clear that the satellite fraction would be systematically too low compared to the correct result (gray area) if the purity correction is not applied. The fraction of satellites for the actual data, once corrected, is in excellent agreement with the expectations from the cosmological simulations. We find a satellite fraction between 20\%-40\% in all redshift bins.

Comparing the result in the three redshift bins, we detect a small, but significant decrease in the satellite fraction $f_{\rm s}$ with redshift at the lower masses (a similar result integrated over all masses was shown in K12).  For the lower masses (i.e., $M \lesssim 10^{10.5} M_\odot$), $f_{\rm s}$ increases systematically from $z \simeq 0.7$ to $z \simeq 0.3$ by up to a factor of two, while at the high mass end there is no significant evolution detected. The increase in the satellite fraction is as expected from the hierarchical buildup of groups over cosmic time.

Comparing our results with that of \cite{vandenbosch2008} from SDSS, we find that the satellite fraction observed by them is systematically lower, by at least 25\%, than the satellite fraction in our lowest redshift bin and decreases more strongly with mass. These differences could be at least partially explained by the fact that \cite{vandenbosch2008} did not correct for impurities in their sample and thus might underestimate the real fraction of satellites (cf.~the dashed line in Figure \ref{fig:satellite_fraction}). Since we used the mock catalogs to classify the galaxies as centrals and satellites, this may reflect a potential bias of the satellite fraction in the mock catalogs (possibly due to the rather high $\sigma_8$ of 0.9 in the Millennium Simulation). While the satellite fraction derived from low-redshift clustering statistics \citep{beutler2013} seems to be consistent with the fraction of \cite{vandenbosch2008}, satellite fractions from galaxy-galaxy weak lensing analyses have too big error bars for resolving this issue \citep{mandelbaum2006}.

\newpage

\subsection{Satellite Quenching Efficiency}\label{sec:epsilon_s}

It has been shown in the low-redshift universe that the red fraction of the centrals $f_{\rm r,c}(M)$ at a given stellar mass $M$ is lower than the red fraction of the corresponding satellite galaxy population $f_{\rm r,s}(M)$ \citep[e.g.,][]{vandenbosch2008,peng2012}.  It should be appreciated that in comparing centrals and satellites at the same stellar mass, the ``centrals'' in question will generally be in different structures (i.e., lower mass halos) than those which contain the ``satellites'', i.e., the central galaxies of the halos containing these particular satellites will generally be of higher stellar mass.  However, our interest is in comparing galaxies of the same stellar mass, so as to remove the (strong) effects of mass-quenching.

Using the red fractions of central and satellite galaxies, we can straightforwardly compute the average ``satellite quenching efficiency'' $\epsilon_{\rm s}$ defined as
\begin{equation}\label{eq:eps_sat}
\epsilon_{\rm s}(M) = \frac{f_{\rm r,s}(M)-f_{\rm r,c}(M)}{f_{\rm b,c}(M)}\:.
\end{equation}
The average in this case is taken over all environments within the halos, i.e., the local densities or group-centric distances. $\epsilon_{\rm s}$ can be interpreted as the fraction of (surviving) blue centrals at a given stellar mass that are quenched when they become satellites by falling into another dark matter halo.  The value as measured in SDSS is about 0.4 (uncorrected for purity) and is essentially constant with mass \citep{vandenbosch2008,peng2012}. However, as mentioned above, the precise value may depend on the specific definition of central and satellite galaxies, and whether one corrects for impurities.

It was argued by \cite{wetzel2012,wetzel2012b} that the calculation of $\epsilon_{\rm s}$ should be based on the red (or blue) fractions of centrals at the time at which the satellites first entered the groups, i.e., when they ceased to be centrals. In their analysis this makes a difference because they adopted a red fraction for centrals that changes rapidly with redshift \cite[see][Figure 3]{wetzel2012b}. We find no evidence in this work for such an evolution back to $z \simeq 0.8$, in general agreement with other investigations in the literature \citep[e.g.,][]{ilbert2013,moustakas2013,hartley2013}. Moreover, the phenomenological model of \cite{peng2010,peng2012} strongly suggests that the red fraction of centrals should be more or less constant for $z \lesssim 1$ because of the observed constant $M^\ast$ in this redshift range \citep[e.g.,][]{ilbert2010,peng2010,ilbert2013}. If there is no strong evolution in $f_{\rm r,c}$ with redshift, then the choice about which epoch to use in calculating $\epsilon_{\rm s}$ is of no practical significance.

In order to compute $f_{\rm r,c}(M)$ and $f_{\rm r,s}(M)$ in the zCOSMOS sample, we again correct for impurities. If $\tilde f_{\rm r,c}$ and $\tilde f_{\rm r,s}$ are the observed (uncorrected) red fractions of centrals and satellites, in a given stellar mass bin, then these will be related to the true red fractions $f_{\rm r,c}$ and $f_{\rm r,s}$ of centrals and satellites by
\begin{align}
\tilde f_{\rm r,c} &= f_{\rm r,c} P_{\rm c} + (1-P_{\rm c}) f_{\rm r,s},\\
\tilde f_{\rm r,s} &= f_{\rm r,s} P_{\rm s} + (1-P_{\rm s}) f_{\rm r,c}.
\end{align}
These simultaneous equations can be trivially solved to give
\begin{align}
f_{\rm r,c} &= \frac{1}{C} \Big(  \tilde f_{\rm r,s} (1-P_{\rm c}) - \tilde f_{\rm r,c} P_{\rm s} \Big) \label{eq:f_r_c}\\
f_{\rm r,s} &= \frac{1}{C} \Big(  \tilde f_{\rm r,c} (1-P_{\rm s}) - \tilde f_{\rm r,s} P_{\rm c} \label{eq:f_r_s}\Big),
\end{align}
where
\begin{equation}
C = (1-P_{\rm c})(1-P_{\rm s})-P_{\rm c}P_{\rm s}.
\end{equation}

Clearly $f_{\rm r,c}$ and $f_{\rm r,s}$ are not well defined if both $P_{\rm c}$ and $P_{\rm s}$ are equal to 50\%, as $C$ is zero in this case, and there is no way to recover the correct red fractions if both the sample of centrals and the sample of satellites are equal mixtures of (real) central and (real) satellites. It should be noted that the above equations are general and may be applied for subsets of centrals and satellites, chosen to maximize their purities, and not just to the dichotomous sample in which every galaxy is retained, provided that the construction of the subsample(s) is not dependent on the color of the galaxy. From this point on, all quoted red fractions are corrected in this way.  

\begin{figure}
	\centering
	\includegraphics[width=0.48\textwidth]{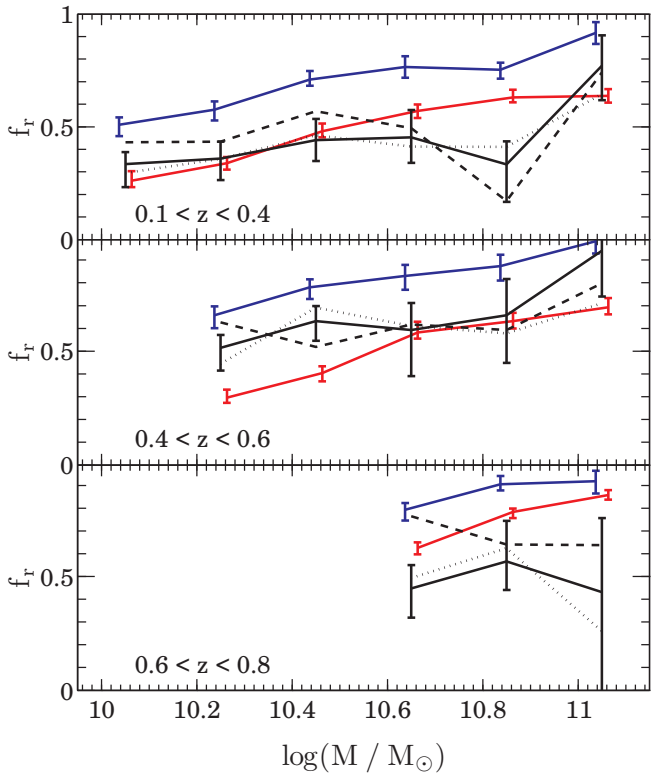}
	\caption{Red fraction of centrals and satellites as a function of stellar mass $M$ within the three redshift bins. The red lines refer to the centrals and the blue lines to the satellites, respectively, of a dichotomous sample. The solid black line corresponds to the satellite quenching efficiency $\epsilon_{\rm s}$ derived from these red fractions. For comparison, the dashed black lines show $\epsilon_{\rm s}$ for high purity samples of centrals and satellites. At any redshift the red fractions of centrals and satellites are clearly distinct and the satellite quenching efficiencies are essentially independent of mass. The dotted black lines show $\epsilon_{\rm s}$, if the galaxy population is split using a color-color diagram instead of a color-mass diagram. All error bars represent the upper and lower quartiles derived from bootstrapping within each mass bin. All quantities are corrected for color incompleteness and impurities due to misidentifications of centrals and satellites.\label{fig:red_fraction}}
\end{figure}

The (corrected) red fractions of centrals and satellites for the dichotomous sample are shown in Figure \ref{fig:red_fraction} for all three redshift bins. The red fractions of the satellites $f_{\rm r,s}$ (blue lines) are always higher than those of the centrals, at all redshifts and at all masses probed by this study. This result is in general agreement with previous studies in the literature \citep{cooper2010,iovino2010,mcgee2011,george2011,presotto2012}, although \cite{iovino2010} could not see a difference between the blue fraction of group galaxies and isolated galaxies at high masses, because they did not differentiate between centrals and satellites and because they are dominated by small groups. Thus, at the high mass end most of the galaxies in their groups will be centrals, as of course most of the isolated objects. As pointed out by \cite{peng2012}, centrals do not exhibit environmental effects.

The corresponding satellite quenching efficiency $\epsilon_{\rm s}$ (black solid lines) is also found to be essentially independent of mass, at each redshift, and to scatter around a value of $\sim\! 0.5$. Within our sample there is no clear change in this quantity with redshift. The value is consistent with that found at low redshift in SDSS \citep{vandenbosch2008,peng2012} especially if the SDSS measurements are corrected (as they should be) for impurities in the samples.

To test our results for robustness, we recomputed $\epsilon_{\rm s}$ for the (corrected) red fractions of high purity samples of centrals and satellites. A high purity subsample of centrals is just given by the set of spectroscopic galaxies that are not associated to any group (i.e., with membership probability $p = 0$), and a high purity subsample of satellites is given by selecting all galaxies with $p > 0.5$, $p_{\rm M} < 0.1$, $p_{\rm MA}<0.1$ that are in groups with at least four observed spectroscopic members.   Both of these samples are about 2\%-7\% purer than the corresponding dichotomous samples, where the difference in purity increases with mass. The gain in purity comes at a cost in completeness, and these samples are about 20\%-50\% smaller than the dichotomous sample depending on mass and redshift. The resulting $\epsilon_{\rm s}$ (dashed black lines) are quite similar to those from the dichotomous sample, within the statistical errors, and thus support the robustness of our results. Only in the lowest mass bin at high redshift is the difference of order of 2$\sigma$.  We also checked that the mass-independence of $\epsilon_{\rm s}$ is not sensitive to our corrections (either for the color completeness or for impurities), as the corresponding $\epsilon_{\rm s}$ derived from the uncorrected red fractions is just systematically shifted down, albeit by about 60\%.

To study the impact of dust-reddened star forming galaxies, we also computed $\epsilon_{\rm s}$ for the case, in which the galaxy population is split using a color-color diagram instead of a color-mass diagram. The resulting $\epsilon_{\rm s}$ (dotted black line) is essentially indistinguishable from the $\epsilon_{\rm s}$ that was derived using the color-mass diagram (solid black line), which indicates that the red centrals and red satellites are similarly contaminated by dust-reddened star forming galaxies. Quantifying the impact from a split using SFR is more difficult, since there is no easy way to correct for the incompleteness of the subpopulations with mass. An approximate calculation, however, shows while in the lowest redshift bin the resulting $\epsilon_{\rm s}$ is basically unchanged, the use of SFR leads to a lower value of $\epsilon_{\rm s}$ of about 0.1-0.2 at higher redshift. We also note that the apparent deviation from mass-independence at the very highest stellar masses, which may also be present in the SDSS analyses of \cite{vandenbosch2008} and \cite{peng2012}, may be an artifact of the contamination from dust-reddened star forming galaxies, since this deviation is mitigated if galaxies are split by means of the color-color diagram (dotted black line).

We conclude that our results are consistent with an essentially mass independent and redshift independent satellite quenching efficiency $\epsilon_{\rm s}$, with a value of about 0.5 (with a scatter of this value from redshift bin to redshift bin of $\lesssim 0.1$), and that $\epsilon_{\rm s}$ is not substantially affected by systematics. It should also be noted that our mean value of 0.5 may well be consistent with the measurements in the low-redshift universe that have yielded a measured value of 0.4, if impurities of the order 10\% are present in the corresponding low-redshift samples of centrals and satellites.

The current analysis therefore extends the previous results about the (average) quenching of satellites, reported in SDSS, back to the epoch when the universe was about a half of its present age.  It is quite striking that the probability of a satellite to have been quenched (because it is a satellite) is the same over a wide range of cosmic epochs, in addition to it apparently being independent of the stellar mass.  This must provide a clue as to the physical nature of satellite-quenching.

\subsection{Stellar Mass Function}

As pointed out in \cite{peng2010} the stellar mass functions of the red and blue galaxy population can give a clear view of quenching processes.  Not least, Peng et al's mass-quenching process is the process which determines the characteristic mass $M^\ast$ of the (surviving) star-forming population, and establishes the $M^\ast$ and $\alpha$ of the passive mass-quenched galaxies in terms of the values of the star-forming population.  

In this section, we therefore compute the mass functions for centrals and satellites split into red and blue galaxies.  In deriving the mass functions, we use the same volume for centrals and satellites, and so the mass functions of the two samples will straightforwardly add to the overall mass function of galaxies in the universe.

\begin{figure*}
	\centering
	\includegraphics[width=0.98\textwidth]{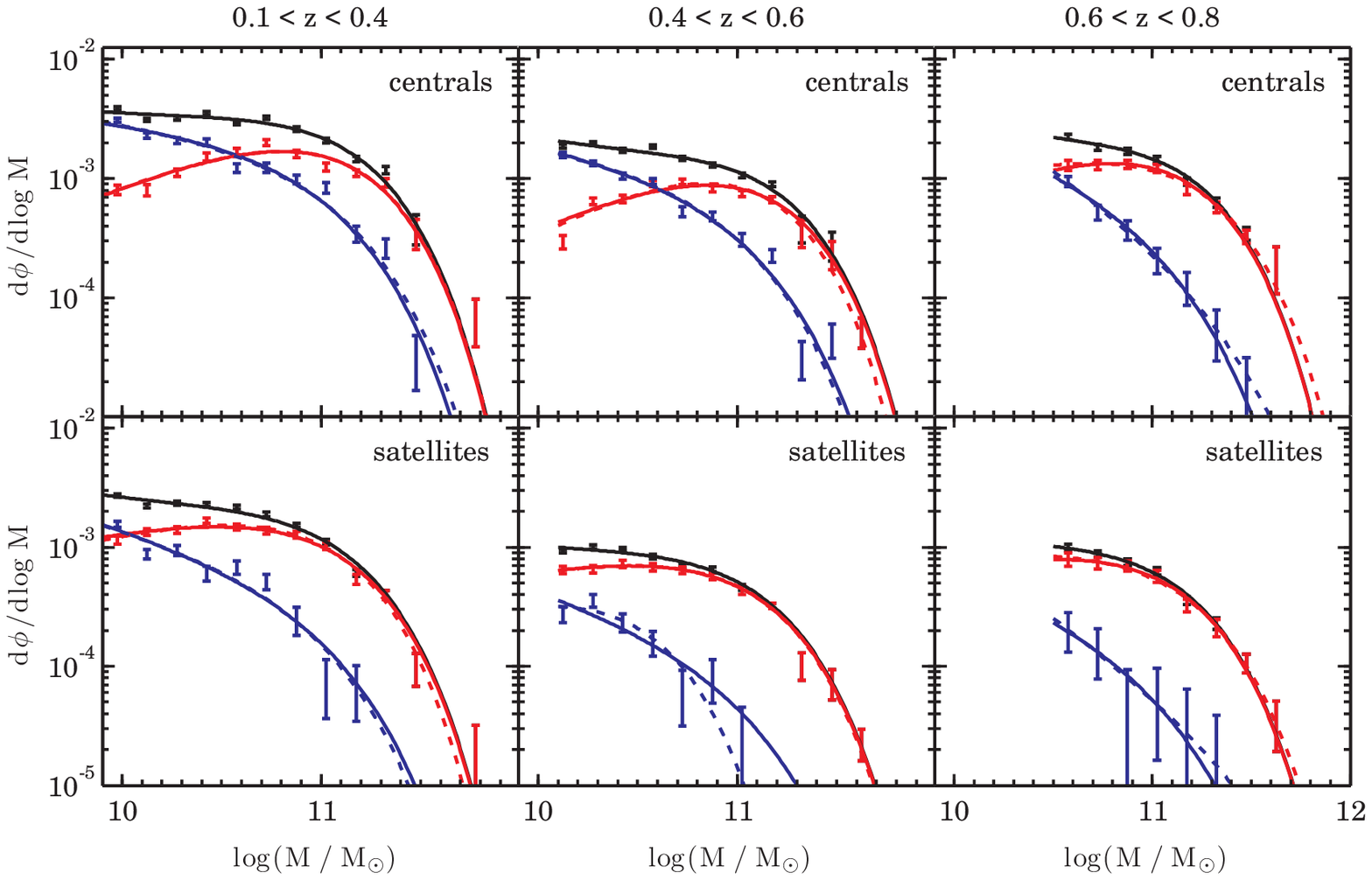}
	\caption{Stellar mass functions for the red and blue populations of centrals and satellites within the three redshift bins. The red points correspond to the red populations, the blue points to the blue. The dashed lines show the best Schechter fits (see Table \ref{tab:schechter_parameters}) and the solid lines the best Schechter fits for a fixed $M^\ast = 10^{10.95}\: M_\odot$ (see the text). The black points and lines correspond to the sum of the blue and red components. All error bars represent the standard deviation derived from bootstrapping.\label{fig:mass_function}}
\end{figure*}

\begin{figure}
	\centering
	\includegraphics[width=0.48\textwidth]{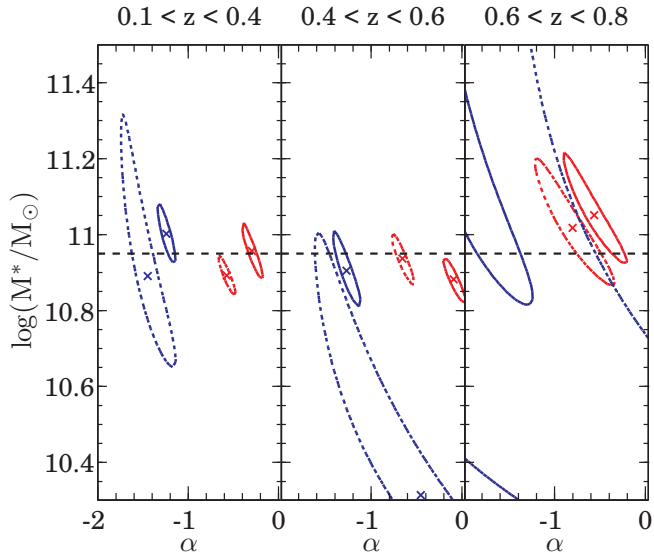}
	\caption{1$\sigma$ confidence regions in the $\alpha$-$M^\ast$ plane from the Schechter fits to the mass functions. The third Schechter parameter $\phi^\ast$ is projected out. The red contours correspond to the red galaxy population and the blue contours to the blue one. The solid contours refer to centrals and the dashed ones to satellites. The crosses mark the best-fitting parameters in the plane. All confidence regions at all redshifts are essentially consistent with the same $M^\ast \simeq 10^{10.95} \: M_\odot$ indicated by the dashed line.\label{fig:contours}}
\end{figure}

To compute the mass functions for the different subpopulations we again correct for the mass-incompleteness in the two color bins, and for the central/satellite-impurities. Similar to Equations (\ref{eq:N_c}) and  (\ref{eq:N_s}), we correct for the impurities as follows
\begin{align}
N_{\rm r,c} &= \tilde N_{\rm r,c} - (1-P_{\rm c})\tilde  N_{\rm c}f_{\rm r,s} + (1-P_{\rm s})\tilde  N_{\rm s}f_{\rm r,c} \\
N_{\rm b,c} &= \tilde  N_{\rm b,c} - (1-P_{\rm c})\tilde  N_{\rm c}f_{\rm b,s} + (1-P_{\rm s})\tilde  N_{\rm s}f_{\rm b,c} \\
N_{\rm r,s} &= \tilde  N_{\rm r,s} - (1-P_{\rm s})\tilde  N_{\rm s}f_{\rm r,c} + (1-P_{\rm c})\tilde  N_{\rm c}f_{\rm r,s} \\
N_{\rm b,s} &= \tilde  N_{\rm b,s} - (1-P_{\rm s})\tilde  N_{\rm s}f_{\rm b,c} + (1-P_{\rm c})\tilde  N_{\rm c}f_{\rm b,s} ,
\end{align}
where the red (and accordingly the blue) fractions are given by Equations (\ref{eq:f_r_c}) and (\ref{eq:f_r_s}). The corrected mass functions for the centrals and satellite galaxies of the red (red points) and blue populations (blue points) in the three redshift bins are shown in Figure \ref{fig:mass_function}. All mass functions can be well fitted by Schechter-functions
\begin{equation}\label{eq:schechter_function}
\frac{d \phi}{d \log M}  = \phi^\ast \left( \frac{M}{M^\ast} \right)^{\alpha+1} \exp \left( -\frac{M}{M^\ast} \right) \ln(10),
\end{equation}
where $\ln = \log_{e}$ denotes the logarithm to the basis $e$, and the best-fitting parameters (corresponding to the dashed curves in Figure \ref{fig:mass_function} and the crosses in Figure \ref{fig:contours}) are summarized in Table \ref{tab:schechter_parameters}. 
\begin{deluxetable}{lccc}
\tablewidth{0pt}
\tablecaption{Best-fitting Schechter Parameters}
\tablehead{
	\colhead{Sample} &
		\colhead{$\phi^\ast$} & 
  \colhead{$\alpha$} &
 	\colhead{$\log M^\ast$} \\
	\colhead{} &
\colhead{$10^{-3}$ Mpc$^{-3}$ dex$^{-1}$} & 
 \colhead{} & 
    \colhead{} 
  }
\startdata
\sidehead{$0.1 < z < 0.4$}
Red centrals &	    $1.89_{-0.19}^{+0.18}$ &  $-0.30_{-0.11}^{+0.12}$ &  $10.96_{-0.07}^{+0.07}$\\
Blue centrals &	    $0.76_{-0.17}^{+0.18}$ &  $-1.24_{-0.09}^{+0.08}$ &  $11.00_{-0.07}^{+0.08}$\\
Red satellites &	    $1.46_{-0.17}^{+0.16}$ &  $-0.57_{-0.09}^{+0.10}$ &  $10.89_{-0.05}^{+0.05}$\\
Blue satellites &    $0.26_{-0.19}^{+0.29}$ &  $-1.45_{-0.29}^{+0.31}$ &  $10.89_{-0.23}^{+0.42}$\\

\hline
\sidehead{$0.4 < z < 0.6$}
Red centrals &     $1.07_{-0.06}^{+0.06}$ &  $-0.09_{-0.11}^{+0.12}$ &   $10.88_{-0.06}^{+0.06}$\\
Blue centrals &    $0.49_{-0.14}^{+0.17}$ &  $-1.27_{-0.14}^{+0.16}$ &   $10.90_{-0.09}^{+0.10}$\\
Red satellites &     $0.62_{-0.09}^{+0.09}$ &  $-0.66_{-0.12}^{+0.13}$ &   $10.94_{-0.07}^{+0.06}$\\
Blue satellites &     $0.33_{-0.27}^{+0.10}$ &  $-0.45_{-1.16}^{+2.63}$ &   $10.31_{-0.52}^{+0.67}$\\

\hline
\sidehead{$0.6 < z < 0.8$}
Red centrals     &    $1.30_{-0.38}^{+0.25}$  &$-0.56_{-0.33}^{+0.33}$   &$11.05_{-0.11}^{+0.16}$\\
Blue centrals   &    $0.09_{-0.08}^{+0.48}$  &$-2.08_{-0.54}^{+0.83}$   &$11.25_{-0.43}^{+0.56}$\\
Red satellites   &    $0.62_{-0.25}^{+0.21}$  &$-0.80_{-0.41}^{+0.46}$   &$11.02_{-0.15}^{+0.18}$\\
Blue satellites  &    $0.01_{-0.00}^{+0.93}$  &$-2.16_{-1.46}^{+5.79}$   &$11.44_{-1.60}^{+0.57}$
\enddata

\tablecomments{The values correspond to the best-fitting parameters of a single Schechter function (\ref{eq:schechter_function}) to the mass functions in Figure \ref{fig:mass_function}, and the error bars indicate their estimated 68\% confidence regions. We want to point out that the error bars of the three parameters for a given fit are strongly correlated (cf.~Figure \ref{fig:contours}).}

\label{tab:schechter_parameters}
\end{deluxetable}

The $1\sigma$ confidence intervals for the Schechter fits in the $M^\ast$-$\alpha$-plane are shown in Figure \ref{fig:contours}, where we projected the three-dimensional confidence region within the $\phi^\ast$-$\alpha$-$M^\ast$ space onto the $\alpha$-$M^\ast$ plane, since for this analysis we are not interested in the normalization of the mass functions.  We clearly see the following trends: first, the error contours increase in area with redshift as the available range of mass shrinks.  Second, the parameters of the red components of both centrals and satellites are typically better constrained than the parameters of the blue mass functions due to the larger numbers of centrals in all samples and the greater curvature of the mass functions due to the less negative $\alpha$. Third, the mass functions of blue satellites are particularly poorly constrained. In fact, at medium and high redshift the information on the blue satellite mass functions is quite limited. 

It is well established that the stellar mass function of the overall galaxy population in the low-redshift universe is a double Schechter function \citep{baldry2006,baldry2008,peng2010,peng2012,baldry2012} and this also extends out to $z \sim 1$ \citep{pozzetti2010,bolzonella2010}.  \cite{peng2010,peng2012} provided a natural explanation for this in terms of the action of their mass-quenching.  The fact that the $M^\ast$ (and $\alpha$) of star-forming galaxies is observed to be constant to at least $z \sim 2$ \citep[e.g.,][]{ilbert2010,peng2010,ilbert2013}, despite the very large increase in stellar mass implied for a galaxy that remains on the star-forming main sequence, places a very strong constraint on the form of mass-quenching (see \citealt{peng2010} for details).   This in turn produces well-defined mass functions of the passive galaxies. 

The approach of \cite{peng2010} predicts precise quantitative relationships between the Schechter parameters of different components of the galaxy population.  Applied to the central/satellite population, as in \cite{peng2012}, these become:
\begin{itemize}

\item The mass functions of blue centrals and blue satellites should be single Schechter functions with the same $M^\ast$ and $\alpha$ and differing only by their normalization $\phi^\ast$.

\item The mass function of red (passive) centrals is produced by mass quenching alone:  it should be a single Schechter function with the same $M^\ast$ as the star-forming galaxies, but with a faint end slope that differs in $\alpha$ by about unity.

\item Both mass quenching and environment quenching act on satellite galaxies.  As a result, the mass function of red (passive) satellites should be a double Schechter function with the same $M^\ast$, but with two different slopes, the one with $\alpha+1$ that is produced by mass quenching (as for the centrals) and another with the same $\alpha$ as the star-forming population that is produced by the environment quenching, which recall is independent of mass. 
\end{itemize}

Thus, all four mass functions are characterized by the same $M^\ast$ and by a faint-end slope that is either $\alpha$ or $\alpha + 1$ (or a superposition of them in the case of the red satellites).  Merging of galaxies can cause small modifications of these relations, but these are evidently small and confined to a small increase in $M^\ast$ of passive central galaxies and, in the main, these relations are observed with impressive precision \citep{peng2012}.  

It is of clear interest to see whether these relations still hold for our set of centrals and satellites at redshifts extending out to unity.  We first simply fit all of the components with a single Schechter function.  It can be seen in Figure \ref{fig:contours} that all the Schechter components are consistent, at all redshifts, with a single value of $M^\ast \simeq 10^{10.95} \: M_\odot$ (dashed line), which in turn is consistent with the values of $M^\ast$ for star forming galaxies observed in COSMOS (see \citealt{peng2010}, Table 1).  It should be noted that the \cite{peng2010,peng2012} values for the SDSS mass functions quote a $\sim\! 0.2$ dex lower value of $M^\ast$ because they were based on calculations of stellar masses that included the effects of mass return from stars.

\begin{figure}
	\centering
	\includegraphics[width=0.48\textwidth]{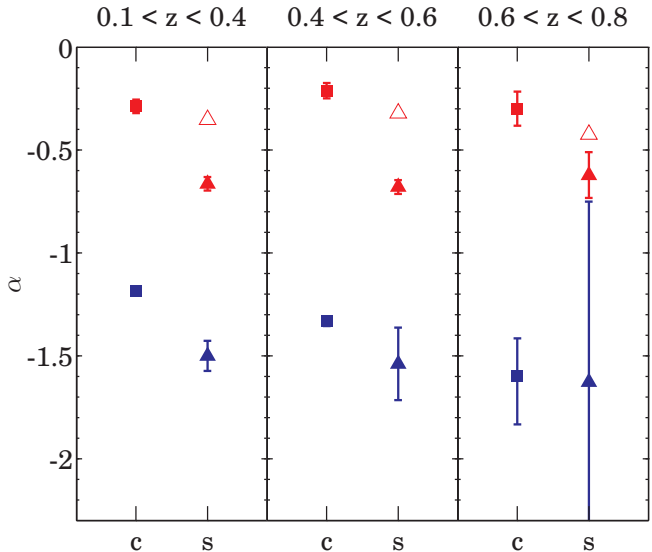}
	\caption{Faint-end slope $\alpha$ if $M^\ast$ is fixed at $10^{10.95} \: M_\odot$. The squares correspond to centrals and the triangles to satellites, where each time the red symbols correspond to the red galaxy population and the blue symbols to the blue one. The filled symbols refer to single Schechter function fits, while the open triangles correspond to double Schechter function fits (see the text). The error bars indicate the standard deviation. For both centrals and satellites the differences in the faint-end slope $\Delta \alpha$ between the red and the blue components are about unity at any redshift.\label{fig:alpha}}
\end{figure}

Since $M^\ast$ is degenerate with $\alpha$, we test the predictions for $\alpha$ from the model of \cite{peng2012} by fixing $M^\ast$ at the value of $10^{10.95} \: M_\odot$, since the model predicts the same $M^\ast$ for all mass functions. The resulting values of $\alpha$ are shown in Figure \ref{fig:alpha} and the corresponding Schechter models are shown in Figure \ref{fig:mass_function} as solid lines. We see a clear difference between the faint-end slope of the red population and the faint-end slope of the blue population, and the difference $\Delta \alpha$ is close to unity in each case at all redshifts.   Unfortunately, in Figure \ref{fig:mass_function} we cannot well isolate the second Schechter component for the red satellites, as it becomes most apparent at $M \lesssim 10^{10} \: M_\odot$ in SDSS.  However, the fact that satellites have a different color distribution to centrals (see Figure \ref{fig:red_fraction}) implies that the effect of the secondary, satellite-quenched, population of passive satellites should be present.  We therefore will assume a double Schechter function for the passive satellites whose components have the same $M^\ast = 10^{10.95} \: M_\odot$, have $\alpha$ differing by unity, and have a relative $\phi^\ast$ that is set by Equation (11) of \cite{peng2012}. This is shown as the second points (open triangles) for the passive satellites in Figure \ref{fig:alpha}.

Fitting then for just the $\alpha$ of each passive population, we find $\alpha \simeq -0.3$ for red centrals and a slightly lower $\alpha$ for red satellites at any redshift. In each case, the blue population mirrors the passive one with a faint end slope $\alpha$ that is steeper by about unity (i.e., $\Delta \alpha \simeq 1$). In the model of \cite{peng2012} the red mass functions are set from the blue ones. Here, we consider the other way around just because the red faint end slopes are better determined observationally. The fact that the $\alpha$ for the satellites is always a bit smaller than the $\alpha$ for the centrals reflects the gradual decrease of the satellite fraction $f_{\rm s}$ with stellar mass of the galaxy (cf.~Figure \ref{fig:satellite_fraction}; see Section 6.2 of \citealt{peng2012} for a discussion of this effect). We checked that our values for $\alpha$ are not sensitive to the specific values of $M_{\rm min}$ (see Table \ref{tab:galaxy_samples}), i.e., changing $M_{\rm min}$ by $\pm 0.2$ dex leads to a shift in $\Delta \alpha$ that is typically of the order of the error bar.

We conclude that in all of our redshift bins, the observed trends are very similar to those described in \cite{peng2012} as observed with SDSS (see their Figure 13 and Table 2), within the obvious limitations of the data, and offer further support for the validity of their simple phenomenological model of the galaxy population, and specifically the role of centrals and satellites out to $z \simeq 0.8$.

\section{Conclusions}
\label{sec:conclusion}

The goal of this paper was to examine the population of central and satellite galaxies at redshifts out to $z \simeq 0.8$ as seen in the extensive group catalog that we have earlier constructed from the zCOSMOS-bright redshift survey. Specifically we have looked at the fraction of centrals and satellites that have been quenched (i.e., in which star formation has effectively ceased) in order to see if the central/satellite paradigm that has been established in the local universe by studies of the SDSS is maintained at these significantly earlier epochs.  Our analysis has been framed by the SDSS analysis of \cite{peng2012}, which itself was developed from the quenching formalism of \cite{peng2010}, and the reader is referred to those papers for detailed discussion.

Our analysis has carefully corrected for both the different mass incompletenesses of blue and red galaxies (calibrated from photo-$z$ data) and for the impurities in the samples of centrals and satellites that enter due to both ambiguities in the identification of the central galaxies and through imperfections in the input group catalog. These impurities are calibrated from comparison with mock samples.  For the analysis of the red fractions we are able to work with both a dichotomous sample, in which all spectroscopic galaxies are classified as either central or satellite, and with subsamples of centrals and satellites that are expected to have higher purity.

We find that the satellite quenching efficiency, which is the probability that a galaxy is quenched because it is a satellite, and which is obtained by comparing the red fractions of centrals and satellites at the same stellar mass, is, as locally, independent of stellar mass (except possibly at the very highest masses).   Furthermore, the strength of the satellite quenching efficiency, averaged over all satellites, is very similar in zCOSMOS, out to $z \simeq 0.8$, as in SDSS locally, at a value of about 0.5.

We also find that the mass functions of red and blue central and satellite galaxies broadly follow the expectations established by the \cite{peng2010,peng2012} model, i.e., within the uncertainties they are consistent with the same characteristic $M^\ast$ and the correct relations between the faint-end slopes $\alpha$ of the various populations.  This confirms the applicability of a universal mass-quenching process applying to both satellite and central galaxies.

Taken together, these results go a long way to confirming the operation of the different quenching channels at $z \simeq 0.8$, i.e., the mass-quenching process applying to all galaxies, and the satellite- (or environment-) quenching applying to just satellite galaxies.  Not least it is quite striking that the probability for a satellite galaxy to be quenched (relative to the case if it were a central) given by the (density-averaged) satellite quenching efficiency $\epsilon_{\rm s}$ is apparently unchanged at $z \simeq 0.8$ as today. Of course the corrections applied to derive these results are significant and precise quantitative values should be treated with caution.

We want to thank the referee for useful comments. This research was supported by the Swiss National Science Foundation, and it is based on observations undertaken at the European Southern Observatory (ESO) Very Large Telescope (VLT) under the Large Program 175.A-0839. The Millennium Simulation databases used in this paper and the web application providing online access to them were constructed as part of the activities of the German Astrophysical Virtual Observatory.

\bibliographystyle{apj3}
\bibliography{apj-jour,bibliography}

\begin{thebibliography}{52}
\expandafter\ifx\csname natexlab\endcsname\relax\def\natexlab#1{#1}\fi
\definecolor{pink}{rgb}{1,0,1}

\bibitem[{{Baldry} {et~al.}(2006){Baldry}, {Balogh}, {Bower}, {Glazebrook},
  {Nichol}, {Bamford}, \& {Budavari}}]{baldry2006}
{Baldry}, I.~K., {Balogh}, M.~L., {Bower}, R.~G., {Glazebrook}, K., {Nichol},
  R.~C., {Bamford}, S.~P., \& {Budavari}, T. 2006,
  \href{http://dx.doi.org/10.1111/j.1365-2966.2006.11081.x}{\textcolor{pink}{\mnras}},
  \href{http://adsabs.harvard.edu/abs/2006MNRAS.373..469B}{\textcolor{blue}{\href{http://adsabs.harvard.edu/abs/2006MNRAS.373..469B}{\textcolor{blue}{373}},
  469}}

\bibitem[{{Baldry} {et~al.}(2008){Baldry}, {Glazebrook}, \&
  {Driver}}]{baldry2008}
{Baldry}, I.~K., {Glazebrook}, K., \& {Driver}, S.~P. 2008,
  \href{http://dx.doi.org/10.1111/j.1365-2966.2008.13348.x}{\textcolor{pink}{\mnras}},
  \href{http://adsabs.harvard.edu/abs/2008MNRAS.388..945B}{\textcolor{blue}{\href{http://adsabs.harvard.edu/abs/2008MNRAS.388..945B}{\textcolor{blue}{388}},
  945}}

\bibitem[{{Baldry} {et~al.}(2012){Baldry}, {Driver}, {Loveday}, {Taylor},
  {Kelvin}, {Liske}, {Norberg}, {Robotham}, {Brough}, {Hopkins}, {Bamford},
  {Peacock}, {Bland-Hawthorn}, {Conselice}, {Croom}, {Jones}, {Parkinson},
  {Popescu}, {Prescott}, {Sharp}, \& {Tuffs}}]{baldry2012}
{Baldry}, I.~K., {et~al.} 2012,
  \href{http://dx.doi.org/10.1111/j.1365-2966.2012.20340.x}{\textcolor{pink}{\mnras}},
  \href{http://adsabs.harvard.edu/abs/2012MNRAS.421..621B}{\textcolor{blue}{\href{http://adsabs.harvard.edu/abs/2012MNRAS.421..621B}{\textcolor{blue}{421}},
  621}}

\bibitem[{{Balogh} {et~al.}(2004){Balogh}, {Baldry}, {Nichol}, {Miller},
  {Bower}, \& {Glazebrook}}]{balogh2004}
{Balogh}, M.~L., {Baldry}, I.~K., {Nichol}, R., {Miller}, C., {Bower}, R., \&
  {Glazebrook}, K. 2004,
  \href{http://dx.doi.org/10.1086/426079}{\textcolor{pink}{\apjl}},
  \href{http://adsabs.harvard.edu/abs/2004ApJ...615L.101B}{\textcolor{blue}{\href{http://adsabs.harvard.edu/abs/2004ApJ...615L.101B}{\textcolor{blue}{615}},
  L101}}

\bibitem[{{Bell} {et~al.}(2004){Bell}, {Wolf}, {Meisenheimer}, {Rix}, {Borch},
  {Dye}, {Kleinheinrich}, {Wisotzki}, \& {McIntosh}}]{bell2004}
{Bell}, E.~F., {et~al.} 2004,
  \href{http://dx.doi.org/10.1086/420778}{\textcolor{pink}{\apj}},
  \href{http://adsabs.harvard.edu/abs/2004ApJ...608..752B}{\textcolor{blue}{\href{http://adsabs.harvard.edu/abs/2004ApJ...608..752B}{\textcolor{blue}{608}},
  752}}

\bibitem[{{Beutler} {et~al.}(2013){Beutler}, {Blake}, {Colless}, {Jones},
  {Staveley-Smith}, {Campbell}, {Parker}, {Saunders}, \&
  {Watson}}]{beutler2013}
{Beutler}, F., {et~al.} 2013,
  \href{http://dx.doi.org/10.1093/mnras/sts637}{\textcolor{pink}{\mnras}},
  \href{http://adsabs.harvard.edu/abs/2013MNRAS.429.3604B}{\textcolor{blue}{\href{http://adsabs.harvard.edu/abs/2013MNRAS.429.3604B}{\textcolor{blue}{429}},
  3604}}

\bibitem[{{Bolzonella} {et~al.}(2010){Bolzonella}, {Kova{\v c}}, {Pozzetti},
  {Zucca}, {Cucciati}, {Lilly}, {Peng}, {Iovino}, {Zamorani}, {Vergani},
  {Tasca}, {Lamareille}, {Oesch}, {Caputi}, {Kampczyk}, {Bardelli}, {Maier},
  {Abbas}, {Knobel}, {Scodeggio}, {Carollo}, {Contini}, {Kneib}, {Le
  F{\`e}vre}, {Mainieri}, {Renzini}, {Bongiorno}, {Coppa}, {de la Torre}, {de
  Ravel}, {Franzetti}, {Garilli}, {Le Borgne}, {Le Brun}, {Mignoli},
  {Pell{\'o}}, {Perez-Montero}, {Ricciardelli}, {Silverman}, {Tanaka},
  {Tresse}, {Bottini}, {Cappi}, {Cassata}, {Cimatti}, {Guzzo}, {Koekemoer},
  {Leauthaud}, {Maccagni}, {Marinoni}, {McCracken}, {Memeo}, {Meneux},
  {Porciani}, {Scaramella}, {Aussel}, {Capak}, {Halliday}, {Ilbert},
  {Kartaltepe}, {Salvato}, {Sanders}, {Scarlata}, {Scoville}, {Taniguchi}, \&
  {Thompson}}]{bolzonella2010}
{Bolzonella}, M., {et~al.} 2010,
  \href{http://dx.doi.org/10.1051/0004-6361/200912801}{\textcolor{pink}{\aap}},
  \href{http://adsabs.harvard.edu/abs/2010A\%26A...524A..76B}{\textcolor{blue}{\href{http://adsabs.harvard.edu/abs/2010A\%26A...524A..76B}{\textcolor{blue}{524}},
  A76}}

\bibitem[{{Brammer} {et~al.}(2011){Brammer}, {Whitaker}, {van Dokkum},
  {Marchesini}, {Franx}, {Kriek}, {Labb{\'e}}, {Lee}, {Muzzin}, {Quadri},
  {Rudnick}, \& {Williams}}]{brammer2011}
{Brammer}, G.~B., {et~al.} 2011,
  \href{http://dx.doi.org/10.1088/0004-637X/739/1/24}{\textcolor{pink}{\apj}},
  \href{http://adsabs.harvard.edu/abs/2011ApJ...739...24B}{\textcolor{blue}{\href{http://adsabs.harvard.edu/abs/2011ApJ...739...24B}{\textcolor{blue}{739}},
  24}}

\bibitem[{{Bruzual} \& {Charlot}(2003)}]{bruzual2003}
{Bruzual}, G., \& {Charlot}, S. 2003,
  \href{http://dx.doi.org/10.1046/j.1365-8711.2003.06897.x}{\textcolor{pink}{\mnras}},
  \href{http://adsabs.harvard.edu/abs/2003MNRAS.344.1000B}{\textcolor{blue}{\href{http://adsabs.harvard.edu/abs/2003MNRAS.344.1000B}{\textcolor{blue}{344}},
  1000}}

\bibitem[{{Bundy} {et~al.}(2010){Bundy}, {Scarlata}, {Carollo}, {Ellis},
  {Drory}, {Hopkins}, {Salvato}, {Leauthaud}, {Koekemoer}, {Murray}, {Ilbert},
  {Oesch}, {Ma}, {Capak}, {Pozzetti}, \& {Scoville}}]{bundy2010}
{Bundy}, K., {et~al.} 2010,
  \href{http://dx.doi.org/10.1088/0004-637X/719/2/1969}{\textcolor{pink}{\apj}},
  \href{http://adsabs.harvard.edu/abs/2010ApJ...719.1969B}{\textcolor{blue}{\href{http://adsabs.harvard.edu/abs/2010ApJ...719.1969B}{\textcolor{blue}{719}},
  1969}}

\bibitem[{{Calzetti} {et~al.}(2000){Calzetti}, {Armus}, {Bohlin}, {Kinney},
  {Koornneef}, \& {Storchi-Bergmann}}]{calzetti2000}
{Calzetti}, D., {Armus}, L., {Bohlin}, R.~C., {Kinney}, A.~L., {Koornneef}, J.,
  \& {Storchi-Bergmann}, T. 2000,
  \href{http://dx.doi.org/10.1086/308692}{\textcolor{pink}{\apj}},
  \href{http://adsabs.harvard.edu/abs/2000ApJ...533..682C}{\textcolor{blue}{\href{http://adsabs.harvard.edu/abs/2000ApJ...533..682C}{\textcolor{blue}{533}},
  682}}

\bibitem[{{Carollo} {et~al.}(2012){Carollo}, {Cibinel}, {Lilly}, {Miniati},
  {Norberg}, {Silverman}, {van Gorkom}, {Cameron}, {Finoguenov}, {Pipino},
  {Rudick}, {Lu}, \& {Peng}}]{carollo2012}
{Carollo}, C.~M., {et~al.} 2012,
  \href{http://arxiv.org/abs/1206.5807}{\textcolor{black}{arXiv:}\textcolor{blue}{1206.5807
  [astro-ph.CO]}}

\bibitem[{{Chabrier}(2003)}]{chabrier2003}
{Chabrier}, G. 2003,
  \href{http://dx.doi.org/10.1086/376392}{\textcolor{pink}{\pasp}},
  \href{http://adsabs.harvard.edu/abs/2003PASP..115..763C}{\textcolor{blue}{\href{http://adsabs.harvard.edu/abs/2003PASP..115..763C}{\textcolor{blue}{115}},
  763}}

\bibitem[{{Cheung} {et~al.}(2012){Cheung}, {Faber}, {Koo}, {Dutton}, {Simard},
  {McGrath}, {Huang}, {Bell}, {Dekel}, {Fang}, {Salim}, {Barro}, {Bundy},
  {Coil}, {Cooper}, {Conselice}, {Davis}, {Dom{\'{\i}}nguez}, {Kassin},
  {Kocevski}, {Koekemoer}, {Lin}, {Lotz}, {Newman}, {Phillips}, {Rosario},
  {Weiner}, \& {Willmer}}]{cheung2012}
{Cheung}, E., {et~al.} 2012,
  \href{http://dx.doi.org/10.1088/0004-637X/760/2/131}{\textcolor{pink}{\apj}},
  \href{http://adsabs.harvard.edu/abs/2012ApJ...760..131C}{\textcolor{blue}{\href{http://adsabs.harvard.edu/abs/2012ApJ...760..131C}{\textcolor{blue}{760}},
  131}}

\bibitem[{{Cibinel} {et~al.}(2012){Cibinel}, {Carollo}, {Lilly}, {Bonoli},
  {Miniati}, {Pipino}, {Silverman}, {van Gorkom}, {Cameron}, {Finoguenov},
  {Norberg}, {Rudick}, {Lu}, \& {Peng}}]{cibinel2012}
{Cibinel}, A., {et~al.} 2012,
  \href{http://arxiv.org/abs/1206.6496}{\textcolor{black}{arXiv:}\textcolor{blue}{1206.6496
  [astro-ph.CO]}}

\bibitem[{{Cirasuolo} {et~al.}(2007){Cirasuolo}, {McLure}, {Dunlop}, {Almaini},
  {Foucaud}, {Smail}, {Sekiguchi}, {Simpson}, {Eales}, {Dye}, {Watson}, {Page},
  \& {Hirst}}]{cirasuolo2007}
{Cirasuolo}, M., {et~al.} 2007,
  \href{http://dx.doi.org/10.1111/j.1365-2966.2007.12038.x}{\textcolor{pink}{\mnras}},
  \href{http://adsabs.harvard.edu/abs/2007MNRAS.380..585C}{\textcolor{blue}{\href{http://adsabs.harvard.edu/abs/2007MNRAS.380..585C}{\textcolor{blue}{380}},
  585}}

\bibitem[{{Cooper} {et~al.}(2010){Cooper}, {Coil}, {Gerke}, {Newman}, {Bundy},
  {Conselice}, {Croton}, {Davis}, {Faber}, {Guhathakurta}, {Koo}, {Lin},
  {Weiner}, {Willmer}, \& {Yan}}]{cooper2010}
{Cooper}, M.~C., {et~al.} 2010,
  \href{http://dx.doi.org/10.1111/j.1365-2966.2010.17312.x}{\textcolor{pink}{\mnras}},
  \href{http://adsabs.harvard.edu/abs/2010MNRAS.409..337C}{\textcolor{blue}{\href{http://adsabs.harvard.edu/abs/2010MNRAS.409..337C}{\textcolor{blue}{409}},
  337}}

\bibitem[{{Croton} {et~al.}(2006){Croton}, {Springel}, {White}, {De Lucia},
  {Frenk}, {Gao}, {Jenkins}, {Kauffmann}, {Navarro}, \& {Yoshida}}]{croton2006}
{Croton}, D.~J., {et~al.} 2006,
  \href{http://dx.doi.org/10.1111/j.1365-2966.2005.09675.x}{\textcolor{pink}{\mnras}},
  \href{http://adsabs.harvard.edu/abs/2006MNRAS.365...11C}{\textcolor{blue}{\href{http://adsabs.harvard.edu/abs/2006MNRAS.365...11C}{\textcolor{blue}{365}},
  11}}

\bibitem[{{Drory} {et~al.}(2009){Drory}, {Bundy}, {Leauthaud}, {Scoville},
  {Capak}, {Ilbert}, {Kartaltepe}, {Kneib}, {McCracken}, {Salvato}, {Sanders},
  {Thompson}, \& {Willott}}]{drory2009}
{Drory}, N., {et~al.} 2009,
  \href{http://dx.doi.org/10.1088/0004-637X/707/2/1595}{\textcolor{pink}{\apj}},
  \href{http://adsabs.harvard.edu/abs/2009ApJ...707.1595D}{\textcolor{blue}{\href{http://adsabs.harvard.edu/abs/2009ApJ...707.1595D}{\textcolor{blue}{707}},
  1595}}

\bibitem[{{George} {et~al.}(2011){George}, {Leauthaud}, {Bundy}, {Finoguenov},
  {Tinker}, {Lin}, {Mei}, {Kneib}, {Aussel}, {Behroozi}, {Busha}, {Capak},
  {Coccato}, {Covone}, {Faure}, {Fiorenza}, {Ilbert}, {Le Floc'h}, {Koekemoer},
  {Tanaka}, {Wechsler}, \& {Wolk}}]{george2011}
{George}, M.~R., {et~al.} 2011,
  \href{http://dx.doi.org/10.1088/0004-637X/742/2/125}{\textcolor{pink}{\apj}},
  \href{http://adsabs.harvard.edu/abs/2011ApJ...742..125G}{\textcolor{blue}{\href{http://adsabs.harvard.edu/abs/2011ApJ...742..125G}{\textcolor{blue}{742}},
  125}}

\bibitem[{{Guo} {et~al.}(2011){Guo}, {White}, {Boylan-Kolchin}, {De Lucia},
  {Kauffmann}, {Lemson}, {Li}, {Springel}, \& {Weinmann}}]{guo2011}
{Guo}, Q., {et~al.} 2011,
  \href{http://dx.doi.org/10.1111/j.1365-2966.2010.18114.x}{\textcolor{pink}{\mnras}},
  \href{http://adsabs.harvard.edu/abs/2011MNRAS.413..101G}{\textcolor{blue}{\href{http://adsabs.harvard.edu/abs/2011MNRAS.413..101G}{\textcolor{blue}{413}},
  101}}

\bibitem[{{Hartley} {et~al.}(2013){Hartley}, {Almaini}, {Mortlock},
  {Conselice}, {Gr{\"u}tzbauch}, {Simpson}, {Bradshaw}, {Chuter}, {Foucaud},
  {Cirasuolo}, {Dunlop}, {McLure}, \& {Pearce}}]{hartley2013}
{Hartley}, W.~G., {et~al.} 2013,
  \href{http://adsabs.harvard.edu/abs/2013MNRAS.tmp.1080H}{\textcolor{blue}{\href{http://dx.doi.org/10.1093/mnras/stt383}{\textcolor{pink}{\mnras}}}}

\bibitem[{{Henriques} {et~al.}(2012){Henriques}, {White}, {Lemson}, {Thomas},
  {Guo}, {Marleau}, \& {Overzier}}]{henriques2012}
{Henriques}, B.~M.~B., {White}, S.~D.~M., {Lemson}, G., {Thomas}, P.~A., {Guo},
  Q., {Marleau}, G.-D., \& {Overzier}, R.~A. 2012,
  \href{http://dx.doi.org/10.1111/j.1365-2966.2012.20521.x}{\textcolor{pink}{\mnras}},
  \href{http://adsabs.harvard.edu/abs/2012MNRAS.421.2904H}{\textcolor{blue}{\href{http://adsabs.harvard.edu/abs/2012MNRAS.421.2904H}{\textcolor{blue}{421}},
  2904}}

\bibitem[{{Ilbert} {et~al.}(2010){Ilbert}, {Salvato}, {Le Floc'h}, {Aussel},
  {Capak}, {McCracken}, {Mobasher}, {Kartaltepe}, {Scoville}, {Sanders},
  {Arnouts}, {Bundy}, {Cassata}, {Kneib}, {Koekemoer}, {Le F{\`e}vre}, {Lilly},
  {Surace}, {Taniguchi}, {Tasca}, {Thompson}, {Tresse}, {Zamojski}, {Zamorani},
  \& {Zucca}}]{ilbert2010}
{Ilbert}, O., {et~al.} 2010,
  \href{http://dx.doi.org/10.1088/0004-637X/709/2/644}{\textcolor{pink}{\apj}},
  \href{http://adsabs.harvard.edu/abs/2010ApJ...709..644I}{\textcolor{blue}{\href{http://adsabs.harvard.edu/abs/2010ApJ...709..644I}{\textcolor{blue}{709}},
  644}}

\bibitem[{{Ilbert} {et~al.}(2013){Ilbert}, {McCracken}, {Le Fevre}, {Capak},
  {Dunlop}, {Arnouts}, {Aussel}, {Caputi}, {Comparat}, {Guo}, {Hudelot},
  {Kartaltepe}, {Kneib}, {Krogager}, {Le Floc'h}, {Lilly}, {Mellier},
  {Milvang-Jensen}, {Moutard}, {Onodera}, {Renzini}, {Richard}, {Salvato},
  {Sanders}, {Scoville}, {Silverman}, {Taniguchi}, {Tasca}, {Thomas}, {Toft},
  {Tresse}, {Vergani}, {Wolk}, \& {Zirm}}]{ilbert2013}
---. 2013,
  \href{http://arxiv.org/abs/1301.3157}{\textcolor{black}{arXiv:}\textcolor{blue}{1301.3157
  [astro-ph.CO]}}

\bibitem[{{Iovino} {et~al.}(2010){Iovino}, {Cucciati}, {Scodeggio}, {Knobel},
  {Kova{\v c}}, {Lilly}, {Bolzonella}, {Tasca}, {Zamorani}, {Zucca}, {Caputi},
  {Pozzetti}, {Oesch}, {Lamareille}, {Halliday}, {Bardelli}, {Finoguenov},
  {Guzzo}, {Kampczyk}, {Maier}, {Tanaka}, {Vergani}, {Carollo}, {Contini},
  {Kneib}, {Le F{\`e}vre}, {Mainieri}, {Renzini}, {Bongiorno}, {Coppa}, {de la
  Torre}, {de Ravel}, {Franzetti}, {Garilli}, {Le Borgne}, {Le Brun},
  {Mignoli}, {Pell{\`o}}, {Peng}, {Perez-Montero}, {Ricciardelli}, {Silverman},
  {Tresse}, {Abbas}, {Bottini}, {Cappi}, {Cassata}, {Cimatti}, {Koekemoer},
  {Leauthaud}, {Maccagni}, {Marinoni}, {McCracken}, {Memeo}, {Meneux},
  {Porciani}, {Scaramella}, {Schiminovich}, \& {Scoville}}]{iovino2010}
{Iovino}, A., {et~al.} 2010,
  \href{http://dx.doi.org/10.1051/0004-6361/200912558}{\textcolor{pink}{\aap}},
  \href{http://adsabs.harvard.edu/abs/2010A\%26A...509A..40I}{\textcolor{blue}{\href{http://adsabs.harvard.edu/abs/2010A\%26A...509A..40I}{\textcolor{blue}{509}},
  A40}}

\bibitem[{{Kauffmann} {et~al.}(2004){Kauffmann}, {White}, {Heckman},
  {M{\'e}nard}, {Brinchmann}, {Charlot}, {Tremonti}, \&
  {Brinkmann}}]{kauffmann2004}
{Kauffmann}, G., {White}, S.~D.~M., {Heckman}, T.~M., {M{\'e}nard}, B.,
  {Brinchmann}, J., {Charlot}, S., {Tremonti}, C., \& {Brinkmann}, J. 2004,
  \href{http://dx.doi.org/10.1111/j.1365-2966.2004.08117.x}{\textcolor{pink}{\mnras}},
  \href{http://adsabs.harvard.edu/abs/2004MNRAS.353..713K}{\textcolor{blue}{\href{http://adsabs.harvard.edu/abs/2004MNRAS.353..713K}{\textcolor{blue}{353}},
  713}}

\bibitem[{{Kimm} {et~al.}(2009){Kimm}, {Somerville}, {Yi}, {van den Bosch},
  {Salim}, {Fontanot}, {Monaco}, {Mo}, {Pasquali}, {Rich}, \&
  {Yang}}]{kimm2009}
{Kimm}, T., {et~al.} 2009,
  \href{http://dx.doi.org/10.1111/j.1365-2966.2009.14414.x}{\textcolor{pink}{\mnras}},
  \href{http://adsabs.harvard.edu/abs/2009MNRAS.394.1131K}{\textcolor{blue}{\href{http://adsabs.harvard.edu/abs/2009MNRAS.394.1131K}{\textcolor{blue}{394}},
  1131}}

\bibitem[{{Kitzbichler} \& {White}(2007)}]{kitzbichler2007}
{Kitzbichler}, M.~G., \& {White}, S.~D.~M. 2007,
  \href{http://dx.doi.org/10.1111/j.1365-2966.2007.11458.x}{\textcolor{pink}{\mnras}},
  \href{http://adsabs.harvard.edu/abs/2007MNRAS.376....2K}{\textcolor{blue}{\href{http://adsabs.harvard.edu/abs/2007MNRAS.376....2K}{\textcolor{blue}{376}},
  2}}

\bibitem[{{Knobel} {et~al.}(2009){Knobel}, {Lilly}, {Iovino}, {Porciani},
  {Kova{\v c}}, {Cucciati}, {Finoguenov}, {Kitzbichler}, {Carollo}, {Contini},
  {Kneib}, {Le F{\`e}vre}, {Mainieri}, {Renzini}, {Scodeggio}, {Zamorani},
  {Bardelli}, {Bolzonella}, {Bongiorno}, {Caputi}, {Coppa}, {de la Torre}, {de
  Ravel}, {Franzetti}, {Garilli}, {Kampczyk}, {Lamareille}, {Le Borgne}, {Le
  Brun}, {Maier}, {Mignoli}, {Pello}, {Peng}, {Perez Montero}, {Ricciardelli},
  {Silverman}, {Tanaka}, {Tasca}, {Tresse}, {Vergani}, {Zucca}, {Abbas},
  {Bottini}, {Cappi}, {Cassata}, {Cimatti}, {Fumana}, {Guzzo}, {Koekemoer},
  {Leauthaud}, {Maccagni}, {Marinoni}, {McCracken}, {Memeo}, {Meneux}, {Oesch},
  {Pozzetti}, \& {Scaramella}}]{knobel2009}
{Knobel}, C., {et~al.} 2009,
  \href{http://dx.doi.org/10.1088/0004-637X/697/2/1842}{\textcolor{pink}{\apj}},
  \href{http://adsabs.harvard.edu/abs/2009ApJ...697.1842K}{\textcolor{blue}{\href{http://adsabs.harvard.edu/abs/2009ApJ...697.1842K}{\textcolor{blue}{697}},
  1842}}

\bibitem[{{Knobel} {et~al.}(2012{\natexlab{a}}){Knobel}, {Lilly}, {Carollo},
  {Contini}, {Kneib}, {Le Fevre}, {Mainieri}, {Renzini}, {Scodeggio},
  {Zamorani}, {Bardelli}, {Bolzonella}, {Bongiorno}, {Caputi}, {Cucciati}, {de
  la Torre}, {de Ravel}, {Franzetti}, {Garilli}, {Iovino}, {Kampczyk}, {Kova{\v
  c}}, {Lamareille}, {Le Borgne}, {Le Brun}, {Maier}, {Mignoli}, {Pello},
  {Peng}, {Perez Montero}, {Presotto}, {Silverman}, {Tanaka}, {Tasca},
  {Tresse}, {Vergani}, {Zucca}, {Barnes}, {Bordoloi}, {Cappi}, {Cimatti},
  {Coppa}, {Koekemoer}, {L{\'o}pez-Sanjuan}, {McCracken}, {Moresco}, {Nair},
  {Pozzetti}, \& {Welikala}}]{knobel2012b}
---. 2012{\natexlab{a}},
  \href{http://dx.doi.org/10.1088/0004-637X/755/1/48}{\textcolor{pink}{\apj}},
  \href{http://adsabs.harvard.edu/abs/2012ApJ...755...48K}{\textcolor{blue}{\href{http://adsabs.harvard.edu/abs/2012ApJ...755...48K}{\textcolor{blue}{755}},
  48}}

\bibitem[{{Knobel} {et~al.}(2012{\natexlab{b}}){Knobel}, {Lilly}, {Iovino},
  {Kova{\v c}}, {Bschorr}, {Presotto}, {Oesch}, {Kampczyk}, {Carollo},
  {Contini}, {Kneib}, {Le Fevre}, {Mainieri}, {Renzini}, {Scodeggio},
  {Zamorani}, {Bardelli}, {Bolzonella}, {Bongiorno}, {Caputi}, {Cucciati}, {de
  la Torre}, {de Ravel}, {Franzetti}, {Garilli}, {Lamareille}, {Le Borgne}, {Le
  Brun}, {Maier}, {Mignoli}, {Pello}, {Peng}, {Perez Montero}, {Silverman},
  {Tanaka}, {Tasca}, {Tresse}, {Vergani}, {Zucca}, {Barnes}, {Bordoloi},
  {Cappi}, {Cimatti}, {Coppa}, {Koekemoer}, {L{\'o}pez-Sanjuan}, {McCracken},
  {Moresco}, {Nair}, {Pozzetti}, \& {Welikala}}]{knobel2012}
---. 2012{\natexlab{b}},
  \href{http://dx.doi.org/10.1088/0004-637X/753/2/121}{\textcolor{pink}{\apj}},
  \href{http://adsabs.harvard.edu/abs/2012ApJ...753..121K}{\textcolor{blue}{\href{http://adsabs.harvard.edu/abs/2012ApJ...753..121K}{\textcolor{blue}{753}},
  121}}

\bibitem[{{Lilly} {et~al.}(2007){Lilly}, {Le F{\`e}vre}, {Renzini}, {Zamorani},
  {Scodeggio}, {Contini}, {Carollo}, {Hasinger}, {Kneib}, {Iovino}, {Le Brun},
  {Maier}, {Mainieri}, {Mignoli}, {Silverman}, {Tasca}, {Bolzonella},
  {Bongiorno}, {Bottini}, {Capak}, {Caputi}, {Cimatti}, {Cucciati}, {Daddi},
  {Feldmann}, {Franzetti}, {Garilli}, {Guzzo}, {Ilbert}, {Kampczyk}, {Kovac},
  {Lamareille}, {Leauthaud}, {Borgne}, {McCracken}, {Marinoni}, {Pello},
  {Ricciardelli}, {Scarlata}, {Vergani}, {Sanders}, {Schinnerer}, {Scoville},
  {Taniguchi}, {Arnouts}, {Aussel}, {Bardelli}, {Brusa}, {Cappi}, {Ciliegi},
  {Finoguenov}, {Foucaud}, {Franceschini}, {Halliday}, {Impey}, {Knobel},
  {Koekemoer}, {Kurk}, {Maccagni}, {Maddox}, {Marano}, {Marconi}, {Meneux},
  {Mobasher}, {Moreau}, {Peacock}, {Porciani}, {Pozzetti}, {Scaramella},
  {Schiminovich}, {Shopbell}, {Smail}, {Thompson}, {Tresse}, {Vettolani},
  {Zanichelli}, \& {Zucca}}]{lilly2007}
{Lilly}, S.~J., {et~al.} 2007,
  \href{http://dx.doi.org/10.1086/516589}{\textcolor{pink}{\apjs}},
  \href{http://adsabs.harvard.edu/abs/2007ApJS..172...70L}{\textcolor{blue}{\href{http://adsabs.harvard.edu/abs/2007ApJS..172...70L}{\textcolor{blue}{172}},
  70}}

\bibitem[{{Lilly} {et~al.}(2009){Lilly}, {Le Brun}, {Maier}, {Mainieri},
  {Mignoli}, {Scodeggio}, {Zamorani}, {Carollo}, {Contini}, {Kneib}, {Le
  F{\`e}vre}, {Renzini}, {Bardelli}, {Bolzonella}, {Bongiorno}, {Caputi},
  {Coppa}, {Cucciati}, {de la Torre}, {de Ravel}, {Franzetti}, {Garilli},
  {Iovino}, {Kampczyk}, {Kovac}, {Knobel}, {Lamareille}, {Le Borgne}, {Pello},
  {Peng}, {P{\'e}rez-Montero}, {Ricciardelli}, {Silverman}, {Tanaka}, {Tasca},
  {Tresse}, {Vergani}, {Zucca}, {Ilbert}, {Salvato}, {Oesch}, {Abbas},
  {Bottini}, {Capak}, {Cappi}, {Cassata}, {Cimatti}, {Elvis}, {Fumana},
  {Guzzo}, {Hasinger}, {Koekemoer}, {Leauthaud}, {Maccagni}, {Marinoni},
  {McCracken}, {Memeo}, {Meneux}, {Porciani}, {Pozzetti}, {Sanders},
  {Scaramella}, {Scarlata}, {Scoville}, {Shopbell}, \& {Taniguchi}}]{lilly2009}
---. 2009,
  \href{http://dx.doi.org/10.1088/0067-0049/184/2/218}{\textcolor{pink}{\apjs}},
  \href{http://adsabs.harvard.edu/abs/2009ApJS..184..218L}{\textcolor{blue}{\href{http://adsabs.harvard.edu/abs/2009ApJS..184..218L}{\textcolor{blue}{184}},
  218}}

\bibitem[{{Mandelbaum} {et~al.}(2006){Mandelbaum}, {Seljak}, {Kauffmann},
  {Hirata}, \& {Brinkmann}}]{mandelbaum2006}
{Mandelbaum}, R., {Seljak}, U., {Kauffmann}, G., {Hirata}, C.~M., \&
  {Brinkmann}, J. 2006,
  \href{http://dx.doi.org/10.1111/j.1365-2966.2006.10156.x}{\textcolor{pink}{\mnras}},
  \href{http://adsabs.harvard.edu/abs/2006MNRAS.368..715M}{\textcolor{blue}{\href{http://adsabs.harvard.edu/abs/2006MNRAS.368..715M}{\textcolor{blue}{368}},
  715}}

\bibitem[{{McGee} {et~al.}(2011){McGee}, {Balogh}, {Wilman}, {Bower},
  {Mulchaey}, {Parker}, \& {Oemler}}]{mcgee2011}
{McGee}, S.~L., {Balogh}, M.~L., {Wilman}, D.~J., {Bower}, R.~G., {Mulchaey},
  J.~S., {Parker}, L.~C., \& {Oemler}, A. 2011,
  \href{http://dx.doi.org/10.1111/j.1365-2966.2010.18189.x}{\textcolor{pink}{\mnras}},
  \href{http://adsabs.harvard.edu/abs/2011MNRAS.413..996M}{\textcolor{blue}{\href{http://adsabs.harvard.edu/abs/2011MNRAS.413..996M}{\textcolor{blue}{413}},
  996}}

\bibitem[{{Mignoli} {et~al.}(2009){Mignoli}, {Zamorani}, {Scodeggio},
  {Cimatti}, {Halliday}, {Lilly}, {Pozzetti}, {Vergani}, {Carollo}, {Contini},
  {Le F{\'e}vre}, {Mainieri}, {Renzini}, {Bardelli}, {Bolzonella}, {Bongiorno},
  {Caputi}, {Coppa}, {Cucciati}, {de La Torre}, {de Ravel}, {Franzetti},
  {Garilli}, {Iovino}, {Kampczyk}, {Kneib}, {Knobel}, {Kova{\v c}},
  {Lamareille}, {Le Borgne}, {Le Brun}, {Maier}, {Pell{\`o}}, {Peng}, {Perez
  Montero}, {Ricciardelli}, {Scarlata}, {Silverman}, {Tanaka}, {Tasca},
  {Tresse}, {Zucca}, {Abbas}, {Bottini}, {Capak}, {Cappi}, {Cassata}, {Fumana},
  {Guzzo}, {Leauthaud}, {Maccagni}, {Marinoni}, {McCracken}, {Memeo}, {Meneux},
  {Oesch}, {Porciani}, {Scaramella}, \& {Scoville}}]{mignoli2009}
{Mignoli}, M., {et~al.} 2009,
  \href{http://dx.doi.org/10.1051/0004-6361:200810520}{\textcolor{pink}{\aap}},
  \href{http://adsabs.harvard.edu/abs/2009A\%26A...493...39M}{\textcolor{blue}{\href{http://adsabs.harvard.edu/abs/2009A\%26A...493...39M}{\textcolor{blue}{493}},
  39}}

\bibitem[{{Mok} {et~al.}(2013){Mok}, {Balogh}, {McGee}, {Wilman}, {Finoguenov},
  {Tanaka}, {Giodini}, {Bower}, {Connelly}, {Hou}, {Mulchaey}, \&
  {Parker}}]{mok2013}
{Mok}, A., {et~al.} 2013,
  \href{http://dx.doi.org/10.1093/mnras/stt251}{\textcolor{pink}{\mnras}},
  \href{http://adsabs.harvard.edu/abs/2013MNRAS.431.1090M}{\textcolor{blue}{\href{http://adsabs.harvard.edu/abs/2013MNRAS.431.1090M}{\textcolor{blue}{431}},
  1090}}

\bibitem[{{Moustakas} {et~al.}(2013){Moustakas}, {Coil}, {Aird}, {Blanton},
  {Cool}, {Eisenstein}, {Mendez}, {Wong}, {Zhu}, \& {Arnouts}}]{moustakas2013}
{Moustakas}, J., {et~al.} 2013,
  \href{http://dx.doi.org/10.1088/0004-637X/767/1/50}{\textcolor{pink}{\apj}},
  \href{http://adsabs.harvard.edu/abs/2013ApJ...767...50M}{\textcolor{blue}{\href{http://adsabs.harvard.edu/abs/2013ApJ...767...50M}{\textcolor{blue}{767}},
  50}}

\bibitem[{{Peng} {et~al.}(2012){Peng}, {Lilly}, {Renzini}, \&
  {Carollo}}]{peng2012}
{Peng}, Y.-j., {Lilly}, S.~J., {Renzini}, A., \& {Carollo}, M. 2012,
  \href{http://dx.doi.org/10.1088/0004-637X/757/1/4}{\textcolor{pink}{\apj}},
  \href{http://adsabs.harvard.edu/abs/2012ApJ...757....4P}{\textcolor{blue}{\href{http://adsabs.harvard.edu/abs/2012ApJ...757....4P}{\textcolor{blue}{757}},
  4}}

\bibitem[{{Peng} {et~al.}(2010){Peng}, {Lilly}, {Kova{\v c}}, {Bolzonella},
  {Pozzetti}, {Renzini}, {Zamorani}, {Ilbert}, {Knobel}, {Iovino}, {Maier},
  {Cucciati}, {Tasca}, {Carollo}, {Silverman}, {Kampczyk}, {de Ravel},
  {Sanders}, {Scoville}, {Contini}, {Mainieri}, {Scodeggio}, {Kneib}, {Le
  F{\`e}vre}, {Bardelli}, {Bongiorno}, {Caputi}, {Coppa}, {de la Torre},
  {Franzetti}, {Garilli}, {Lamareille}, {Le Borgne}, {Le Brun}, {Mignoli},
  {Perez Montero}, {Pello}, {Ricciardelli}, {Tanaka}, {Tresse}, {Vergani},
  {Welikala}, {Zucca}, {Oesch}, {Abbas}, {Barnes}, {Bordoloi}, {Bottini},
  {Cappi}, {Cassata}, {Cimatti}, {Fumana}, {Hasinger}, {Koekemoer},
  {Leauthaud}, {Maccagni}, {Marinoni}, {McCracken}, {Memeo}, {Meneux}, {Nair},
  {Porciani}, {Presotto}, \& {Scaramella}}]{peng2010}
{Peng}, Y.-j., {et~al.} 2010,
  \href{http://dx.doi.org/10.1088/0004-637X/721/1/193}{\textcolor{pink}{\apj}},
  \href{http://adsabs.harvard.edu/abs/2010ApJ...721..193P}{\textcolor{blue}{\href{http://adsabs.harvard.edu/abs/2010ApJ...721..193P}{\textcolor{blue}{721}},
  193}}

\bibitem[{{Pozzetti} {et~al.}(2010){Pozzetti}, {Bolzonella}, {Zucca},
  {Zamorani}, {Lilly}, {Renzini}, {Moresco}, {Mignoli}, {Cassata}, {Tasca},
  {Lamareille}, {Maier}, {Meneux}, {Halliday}, {Oesch}, {Vergani}, {Caputi},
  {Kova{\v c}}, {Cimatti}, {Cucciati}, {Iovino}, {Peng}, {Carollo}, {Contini},
  {Kneib}, {Le F{\'e}vre}, {Mainieri}, {Scodeggio}, {Bardelli}, {Bongiorno},
  {Coppa}, {de la Torre}, {de Ravel}, {Franzetti}, {Garilli}, {Kampczyk},
  {Knobel}, {Le Borgne}, {Le Brun}, {Pell{\`o}}, {Perez Montero},
  {Ricciardelli}, {Silverman}, {Tanaka}, {Tresse}, {Abbas}, {Bottini}, {Cappi},
  {Guzzo}, {Koekemoer}, {Leauthaud}, {Maccagni}, {Marinoni}, {McCracken},
  {Memeo}, {Porciani}, {Scaramella}, {Scarlata}, \& {Scoville}}]{pozzetti2010}
{Pozzetti}, L., {et~al.} 2010,
  \href{http://dx.doi.org/10.1051/0004-6361/200913020}{\textcolor{pink}{\aap}},
  \href{http://adsabs.harvard.edu/abs/2010A\%26A...523A..13P}{\textcolor{blue}{\href{http://adsabs.harvard.edu/abs/2010A\%26A...523A..13P}{\textcolor{blue}{523}},
  A13}}

\bibitem[{{Presotto} {et~al.}(2012){Presotto}, {Iovino}, {Scodeggio},
  {Cucciati}, {Knobel}, {Bolzonella}, {Oesch}, {Finoguenov}, {Tanaka}, {Kova{\v
  c}}, {Peng}, {Zamorani}, {Bardelli}, {Pozzetti}, {Kampczyk},
  {L{\'o}pez-Sanjuan}, {Vergani}, {Zucca}, {Tasca}, {Carollo}, {Contini},
  {Kneib}, {Le F{\`e}vre}, {Lilly}, {Mainieri}, {Renzini}, {Bongiorno},
  {Caputi}, {de la Torre}, {de Ravel}, {Franzetti}, {Garilli}, {Lamareille},
  {Le Borgne}, {Le Brun}, {Maier}, {Mignoli}, {Pell{\`o}}, {Perez-Montero},
  {Ricciardelli}, {Silverman}, {Tresse}, {Barnes}, {Bordoloi}, {Cappi},
  {Cimatti}, {Coppa}, {Koekemoer}, {McCracken}, {Moresco}, {Nair}, \&
  {Welikala}}]{presotto2012}
{Presotto}, V., {et~al.} 2012,
  \href{http://dx.doi.org/10.1051/0004-6361/201118293}{\textcolor{pink}{\aap}},
  \href{http://adsabs.harvard.edu/abs/2012A\%26A...539A..55P}{\textcolor{blue}{\href{http://adsabs.harvard.edu/abs/2012A\%26A...539A..55P}{\textcolor{blue}{539}},
  A55}}

\bibitem[{{Skibba} {et~al.}(2011){Skibba}, {van den Bosch}, {Yang}, {More},
  {Mo}, \& {Fontanot}}]{skibba2011}
{Skibba}, R.~A., {van den Bosch}, F.~C., {Yang}, X., {More}, S., {Mo}, H., \&
  {Fontanot}, F. 2011,
  \href{http://dx.doi.org/10.1111/j.1365-2966.2010.17452.x}{\textcolor{pink}{\mnras}},
  \href{http://adsabs.harvard.edu/abs/2011MNRAS.410..417S}{\textcolor{blue}{\href{http://adsabs.harvard.edu/abs/2011MNRAS.410..417S}{\textcolor{blue}{410}},
  417}}

\bibitem[{{Springel} {et~al.}(2005){Springel}, {White}, {Jenkins}, {Frenk},
  {Yoshida}, {Gao}, {Navarro}, {Thacker}, {Croton}, {Helly}, {Peacock}, {Cole},
  {Thomas}, {Couchman}, {Evrard}, {Colberg}, \& {Pearce}}]{springel2005}
{Springel}, V., {et~al.} 2005,
  \href{http://dx.doi.org/10.1038/nature03597}{\textcolor{pink}{\nat}},
  \href{http://adsabs.harvard.edu/abs/2005Natur.435..629S}{\textcolor{blue}{\href{http://adsabs.harvard.edu/abs/2005Natur.435..629S}{\textcolor{blue}{435}},
  629}}

\bibitem[{{Strateva} {et~al.}(2001){Strateva}, {Ivezi{\'c}}, {Knapp},
  {Narayanan}, {Strauss}, {Gunn}, {Lupton}, {Schlegel}, {Bahcall}, {Brinkmann},
  {Brunner}, {Budav{\'a}ri}, {Csabai}, {Castander}, {Doi}, {Fukugita}, {Gy{\H
  o}ry}, {Hamabe}, {Hennessy}, {Ichikawa}, {Kunszt}, {Lamb}, {McKay},
  {Okamura}, {Racusin}, {Sekiguchi}, {Schneider}, {Shimasaku}, \&
  {York}}]{strateva2001}
{Strateva}, I., {et~al.} 2001,
  \href{http://dx.doi.org/10.1086/323301}{\textcolor{pink}{\aj}},
  \href{http://adsabs.harvard.edu/abs/2001AJ....122.1861S}{\textcolor{blue}{\href{http://adsabs.harvard.edu/abs/2001AJ....122.1861S}{\textcolor{blue}{122}},
  1861}}

\bibitem[{{van den Bosch} {et~al.}(2008){van den Bosch}, {Aquino}, {Yang},
  {Mo}, {Pasquali}, {McIntosh}, {Weinmann}, \& {Kang}}]{vandenbosch2008}
{van den Bosch}, F.~C., {Aquino}, D., {Yang}, X., {Mo}, H.~J., {Pasquali}, A.,
  {McIntosh}, D.~H., {Weinmann}, S.~M., \& {Kang}, X. 2008,
  \href{http://dx.doi.org/10.1111/j.1365-2966.2008.13230.x}{\textcolor{pink}{\mnras}},
  \href{http://adsabs.harvard.edu/abs/2008MNRAS.387...79V}{\textcolor{blue}{\href{http://adsabs.harvard.edu/abs/2008MNRAS.387...79V}{\textcolor{blue}{387}},
  79}}

\bibitem[{{Wetzel} {et~al.}(2012){Wetzel}, {Tinker}, \& {Conroy}}]{wetzel2012}
{Wetzel}, A.~R., {Tinker}, J.~L., \& {Conroy}, C. 2012,
  \href{http://dx.doi.org/10.1111/j.1365-2966.2012.21188.x}{\textcolor{pink}{\mnras}},
  \href{http://adsabs.harvard.edu/abs/2012MNRAS.424..232W}{\textcolor{blue}{\href{http://adsabs.harvard.edu/abs/2012MNRAS.424..232W}{\textcolor{blue}{424}},
  232}}

\bibitem[{{Wetzel} {et~al.}(2013){Wetzel}, {Tinker}, {Conroy}, \& {van den
  Bosch}}]{wetzel2012b}
{Wetzel}, A.~R., {Tinker}, J.~L., {Conroy}, C., \& {van den Bosch}, F.~C. 2013,
  \href{http://adsabs.harvard.edu/abs/2013MNRAS.tmp.1130W}{\textcolor{blue}{\href{http://dx.doi.org/10.1093/mnras/stt469}{\textcolor{pink}{\mnras}}}}

\bibitem[{{Williams} {et~al.}(2009){Williams}, {Quadri}, {Franx}, {van Dokkum},
  \& {Labb{\'e}}}]{williams2009}
{Williams}, R.~J., {Quadri}, R.~F., {Franx}, M., {van Dokkum}, P., \&
  {Labb{\'e}}, I. 2009,
  \href{http://dx.doi.org/10.1088/0004-637X/691/2/1879}{\textcolor{pink}{\apj}},
  \href{http://adsabs.harvard.edu/abs/2009ApJ...691.1879W}{\textcolor{blue}{\href{http://adsabs.harvard.edu/abs/2009ApJ...691.1879W}{\textcolor{blue}{691}},
  1879}}

\bibitem[{{Woo} {et~al.}(2013){Woo}, {Dekel}, {Faber}, {Noeske}, {Koo},
  {Gerke}, {Cooper}, {Salim}, {Dutton}, {Newman}, {Weiner}, {Bundy}, {Willmer},
  {Davis}, \& {Yan}}]{woo2013}
{Woo}, J., {et~al.} 2013,
  \href{http://dx.doi.org/10.1093/mnras/sts274}{\textcolor{pink}{\mnras}},
  \href{http://adsabs.harvard.edu/abs/2013MNRAS.428.3306W}{\textcolor{blue}{\href{http://adsabs.harvard.edu/abs/2013MNRAS.428.3306W}{\textcolor{blue}{428}},
  3306}}

\bibitem[{{Wuyts} {et~al.}(2011){Wuyts}, {F{\"o}rster Schreiber}, {van der
  Wel}, {Magnelli}, {Guo}, {Genzel}, {Lutz}, {Aussel}, {Barro}, {Berta},
  {Cava}, {Graci{\'a}-Carpio}, {Hathi}, {Huang}, {Kocevski}, {Koekemoer},
  {Lee}, {Le Floc'h}, {McGrath}, {Nordon}, {Popesso}, {Pozzi}, {Riguccini},
  {Rodighiero}, {Saintonge}, \& {Tacconi}}]{wuyts2011}
{Wuyts}, S., {et~al.} 2011,
  \href{http://dx.doi.org/10.1088/0004-637X/742/2/96}{\textcolor{pink}{\apj}},
  \href{http://adsabs.harvard.edu/abs/2011ApJ...742...96W}{\textcolor{blue}{\href{http://adsabs.harvard.edu/abs/2011ApJ...742...96W}{\textcolor{blue}{742}},
  96}}

\end{thebibliography}

\end{document}